\documentclass[aip,reprint]{revtex4-1}

\usepackage{amsmath}
\usepackage{amssymb}
\usepackage{graphicx}
\usepackage{tabularx}
\usepackage{array}
\usepackage{booktabs}
\usepackage{xcolor} 
\usepackage[utf8]{inputenc}
\usepackage[T1]{fontenc}

\newcolumntype{C}[1]{>{\centering\arraybackslash}m{#1}}
\newcolumntype{Y}{>{\centering\arraybackslash}X}

\draft 

\begin{document}


\title{A Reduced-Order Particle-in-Cell Method with Azimuthal Fourier-Decomposed Fields for Nominally Axisymmetric Plasmas} 



\author{Shaun Andrews}
\email[]{sa15339@my.bristol.ac.uk}
\affiliation{$^1$Independent Research Consultant, Wiltshire, United Kingdom}


\date{\today}

\begin{abstract}
{\color{red}This article may be downloaded for personal use only. Any other use requires prior permission of the author and AIP Publishing. This article appeared in Phys. Plasmas 33, 093903 (2026) and may be found at https://doi.org/10.1063/5.0347124}\\

A reduced-order Particle-in-Cell method is introduced for kinetic simulation of otherwise axisymmetric cylindrical plasmas that exhibit azimuthal instabilities. The method spatially decomposes all field quantities into a small number of azimuthal Fourier modes $m=0,...,N_m,\quad N_m \ll N_\theta$, reducing the costly three-dimensional field solve $\mathcal{O}(N_zN_rN_\theta)$ to a family of decoupled independent two-dimensional problems $\mathcal{O}((N_m+1)N_zN_r)$ on the meridional plane --- one per mode --- while particles continue to move in full three-dimensional space. Fields are reconstructed at particle positions by coherent superposition of these modal contributions, preserving complete azimuthal variation at a fraction of the cost of a conventional three-dimensional simulation. {\color{black}For initial validation, the method is tested against two benchmark problems with restricted axial dimension:} The diocotron instability of a hollow electron annulus across three geometrically distinct configurations, recovering linear growth rates and eigenpotential structures within 7\% of closed-form analytic predictions, and reproducing the non-linear vortex dynamics characteristic of the instability saturation; and a further benchmark against the community-standard Landmark Penning discharge problem recovers the rotating-spoke frequency, radial plasma profiles, and modal energy hierarchy in quantitative agreement with long-time reference simulations, at approximately 640 CPU-hours --- a factor of 46 speed-up compared to the median benchmark cost. {\color{black}The approach represents a step toward addressing} the important gap between computationally prohibitive full three-dimensional kinetic simulation and the physically limited reduced-dimensionality models on which predictive modelling of anomalous transport in magnetised plasma devices currently relies.
\end{abstract}


\maketitle 

\section{Introduction}
The kinetic simulation of non-equilibrium plasmas represents a central challenge in computational plasma physics. In a wide class of physically and technologically important plasma systems --- spanning magnetic confinement fusion, space and astrophysical plasmas, low-temperature processing discharges, and plasma propulsion devices --- the behaviour is governed by the detailed structure of the particle distribution function in full six-dimensional phase space. Kinetic effects including instabilities, anomalous transport, and wave-particle resonances are beyond the reach of fluid or hybrid descriptions. 

Resolving these phenomena self-consistently from first principles, without the closure assumptions and empirical transport coefficients that fluid models require, demands a computational approach that retains the kinetic character of the plasma in its entirety. Particle-in-Cell (PIC) is the definitive method for fully kinetic plasma simulation \cite{birdsall1991}, representing the distribution function as an ensemble of macroparticles advanced under self-consistent electromagnetic fields obtained by solving Poisson's equation (electrostatic) or the full Maxwell system (electromagnetic) on a spatial grid \cite{b:taccognareview}.

One or two-dimensional PIC treatments are often qualitatively inadequate for a wide class of phenomena whose dynamics are intrinsically three-dimensional. In magnetised plasma devices of engineering relevance --- Hall-effect thrusters \cite{hettheory1,hettheory2}, magnetic nozzles \cite{b:andrews2022,b:hepner2020AIAA}, Penning discharges \cite{powis2018scaling} --- the dominant unresolved physics are often azimuthal in character. Instabilities such as the electron cyclotron drift instability (ECDI), two-stream instability, rotating spoke, and drift wave, all require simultaneous resolution of all three spatial dimensions; their non-linear saturation, cross-dimensional coupling, and contribution to anomalous cross-field transport cannot be captured in any lower-dimensional analogue \cite{b:exb}.

In many applications, the primary motivation for capturing plasma instabilities is indeed the influence on macroscopic plasma distributions and global device performance via that resultant anomalous transport. Despite decades of theoretical and experimental investigation, prediction of anomalous transport coefficients from plasma parameters remains elusive, forcing many contemporary PIC or hybrid codes to rely on phenomenological or empirically tuned expressions \cite{Andrews2025MagneticNozzleBohm, sanchezvillar2023ecr, b:andri}. More recently, models derived from first-principle instability theory \cite{lafleur2016anomII, b:andrewsiepc} and data-driven methods \cite{marks2023,Jorns2018HallCollisionFreq} have also been introduced. However, full three-dimensional kinetic simulation remains the fundamental prerequisite for true predictive modelling.

The computational demands are, however, severe. The spatial grid must resolve the Debye length $\lambda_D$ throughout the domain and the time-step must resolve the electron plasma period $\omega_{pe}^{-1}$. Consequently, for a $d$-dimensional system of characteristic size $L$, the computational cost scales approximately as $(L/\lambda_D)^d \omega_{pe} T$, becoming particularly prohibitive in three dimensions. For example, for a Hall-effect thruster at $n_e\sim10^{17}$ m$^{-3}$, {\color{black}a cubic centimetre domain requires $\mathcal{O}(10^9)$ cells and, assuming even a
sparse loading of $\mathcal{O}(1)$ macroparticle per cell, $\mathcal{O}(10^{13})$ particle-field operations for the $\mathcal{O}(10^4)$ time-steps needed per 1~$\mu$s of physical time; realistic particle loadings push this a further one to two orders of magnitude higher.} Three-dimensional PIC simulations of thruster segments have been reported consuming in excess of $10^6$ CPU-hours per run, confined to sub-centimetre domains and sub-microsecond durations \cite{villafana3D,taccogna3d}. Magnetic nozzles have also been recently modelled in three-dimensions \cite{b:difede2021,Vatansever2025CHAOS, b:AA2022}, with higher operating densities and longer ion transit time-scales further extending the computational burden. In both cases, heavy artificial scaling of plasma or geometric parameters is utilised to enable the simulations to be feasible. It is this fundamental tension between three-dimensional kinetic fidelity and finite computational resources that motivates the development of reduced-cost algorithms that preserve the essential physics at tractable expense.

Several strategies have been proposed to reduce this cost. Sparse-grid methods~\cite{ricketson2015picsparse,garrigues2021sparse} replace the standard PIC mesh with a combination of coarser sub-grids, reducing cell count from $\mathcal{O}(N^d)$ to $\mathcal{O}(N\log N^{d-1})$ and demonstrating speed-ups of around $6.5\times$ in Hall-thruster ECDI simulations~\cite{garrigues2022benchmark}; however, they introduce reconstruction errors at steep gradients and it remains unclear whether they correctly capture cross-dimensional instability coupling. 

Implicit PIC formulations~\cite{langdon1983direct,brackbill1982implicit} relax the Debye-length and time-step constraints by treating the field-particle coupling iteratively, permitting larger time-steps and coarser grids; energy-conserving variants further suppress numerical heating~\cite{chen2011energy}. The cost is that implicit schemes introduce artificial damping and dispersion for the short-wavelength modes central to kinetic instabilities such as the ECDI and two-stream modes, risking suppression or under-prediction of growth rates and non-linear saturation relative to a well-resolved explicit simulation.

{\color{black}Lifschitz {et al.}\cite{lifschitz2009} introduced a PIC method that represented the electromagnetic fields as a finite sum of azimuthal Fourier modes on a finite-difference mesh. The work demonstrated that predominantly axisymmetric electromagnetic problems could be modelled using only a small number of azimuthal harmonics, with laser applications typically retaining only the $m=0$ and $m=1$ modes. The formulation, however, was developed purely for relativistic laser-plasma interaction, where the field equations are governed by Maxwell's equations in highly regular, structured computational domains. It therefore does not address the treatment of bounded electrostatic plasmas with conducting boundaries, non-periodic boundary conditions or the geometric complexity encountered in practical plasma devices. Global spectral solvers, notably the analytical time-domain based extension of Lehe {et al.} \cite{lehe2016}, go further and reduce the three-dimensional field problem into two-dimensional spectral solves in the meridional plane, eliminating the numerical dispersion that afflicts finite-difference schemes. The reliance on global Fourier and Hankel transforms, however, makes parallelisation non-trivial and again restricts practical applicability to periodic or otherwise highly regular domains, rendering them poorly suited to the open boundaries, embedded conductors, and localised boundary conditions typical of plasma device geometries. As such, these solvers also have remained limited to the field of laser-plasma interaction.}

Recently, a more radical separability ansatz has been proposed~\cite{reza2022ropicaxaz}, reproduced from a technique originally developed for magnetron spoke modelling~\cite{Revel2016Pseudo3DPIC}. The three-dimensional potential is decomposed as a sum of independent one-dimensional contributions, $\phi(x,y,z)=\chi(x)+\eta(y)+\zeta(z)$, reducing the three-dimensional Poisson solve to a triplet of coupled one-dimensional problems sharing a common macroparticle ensemble. The computational cost is thus reduced to $\mathcal{O}(dN)$. The method has been reported to recover bulk discharge behaviour and macroscopic mode structures in Hall-thruster, Penning and diocotron configurations\cite{Reza2024PenningAxialRadial,Reza2025Quasi3D}, but the separability ansatz has no rigorous basis in the governing equations: the Poisson equation does not admit separable solutions in the presence of a spatially inhomogeneous charge density, there is no small parameter whose limit recovers the exact solution, and the accuracy depends on the region-decomposition coarseness in a problem-dependent manner that provides no a priori convergence guarantee. Instabilities with oblique wavevectors or correlated cross-dimensional structure may therefore be distorted in both growth rate and non-linear saturation. 

Finally, machine-learning approaches --- including neural-network surrogate field solvers \cite{Markidis2021PINNSolvers,AguilarMarkidis2021Cluster}, data-driven preconditioners \cite{Li2023NeuralPreconditioner}, and reduced-order models \cite{FarajiReza2025MLPlasmaPerspective} --- have attracted growing interest but face fundamental constraints in this setting: generating training data for high-resolution three-dimensional PIC problems is itself extremely expensive, purely data-driven surrogates offer no guarantees of conservation or long-term stability, and the risk of out-of-distribution failure is acute in plasma applications where small parameter variations can produce qualitatively different dynamics. These limitations collectively motivate a method that exploits exact mathematical structure of the governing equations rather than approximating or learning it.

{\color{black}Relative to the closest existing azimuthal-mode PIC scheme of Lifschitz et al.\cite{lifschitz2009}, the present work introduces a reduced-order cylindrical electrostatic-magnetostatic PIC algorithm designed for the general simulation of low-temperature plasmas that exhibit nominal axisymmetry but with azimuthal instability.} The method decomposes the electrostatic potential into $m=0,...,N_m$ azimuthal Fourier modes, reducing a three-dimensional field solve to $N_m+1$ independent two-dimensional Poisson equations in the $(r,z)$ plane. These are solved using a finite-volume discretisation and a geometric multigrid algorithm with a mode-adapted preconditioner. Particles are advanced in full three-dimensional coordinates with fields reconstructed from the modal contributions. The method enables direct simulation of instability phenomena and self-consistent computation of transport coefficients --- {\color{black}intended to help address a capability gap} in predictive device modelling.

The formulation is intended for nominally axisymmetric plasma systems whose non-axisymmetric dynamics can be represented by a small number of azimuthal harmonics. The approach does not eliminate the need for fully three-dimensional kinetic simulation in arbitrary geometries, but instead targets the important class of (mostly magnetised) plasma devices whose dominant departures from axisymmetry are low-order azimuthal instabilities. Under these conditions, the modal representation provides a substantial reduction in computational cost while retaining fully three-dimensional particle dynamics. {\color{black}While the approach shares the underlying philosophy of some other reduced-order PIC methods that represent fields spectrally in an ignorable coordinate, the distinguishing feature here is the formulation for cylindrical low-temperature plasmas relevant to plasma propulsion and other E$\times$B devices, permitting efficient treatment of low-order azimuthal instabilities while retaining compatibility and versatility with conventional explicit PIC workflows.}

The remainder of this article is organised as follows. Section 2 derives the azimuthal mode decomposition, the modal Poisson equations, particle deposition-interpolation and field reconstruction. Section 3 describes the numerical implementation, including the finite-volume stencil, multigrid solver, preconditioning strategy, and stability requirements. Section 4 then presents validation against the diocotron instability and a benchmark Penning Discharge case. Section 5 contains the conclusions.



\section{Theoretical Formulation}
\label{sec:theory}

\begin{figure*}[!t]
\centering
\includegraphics[width=\textwidth]{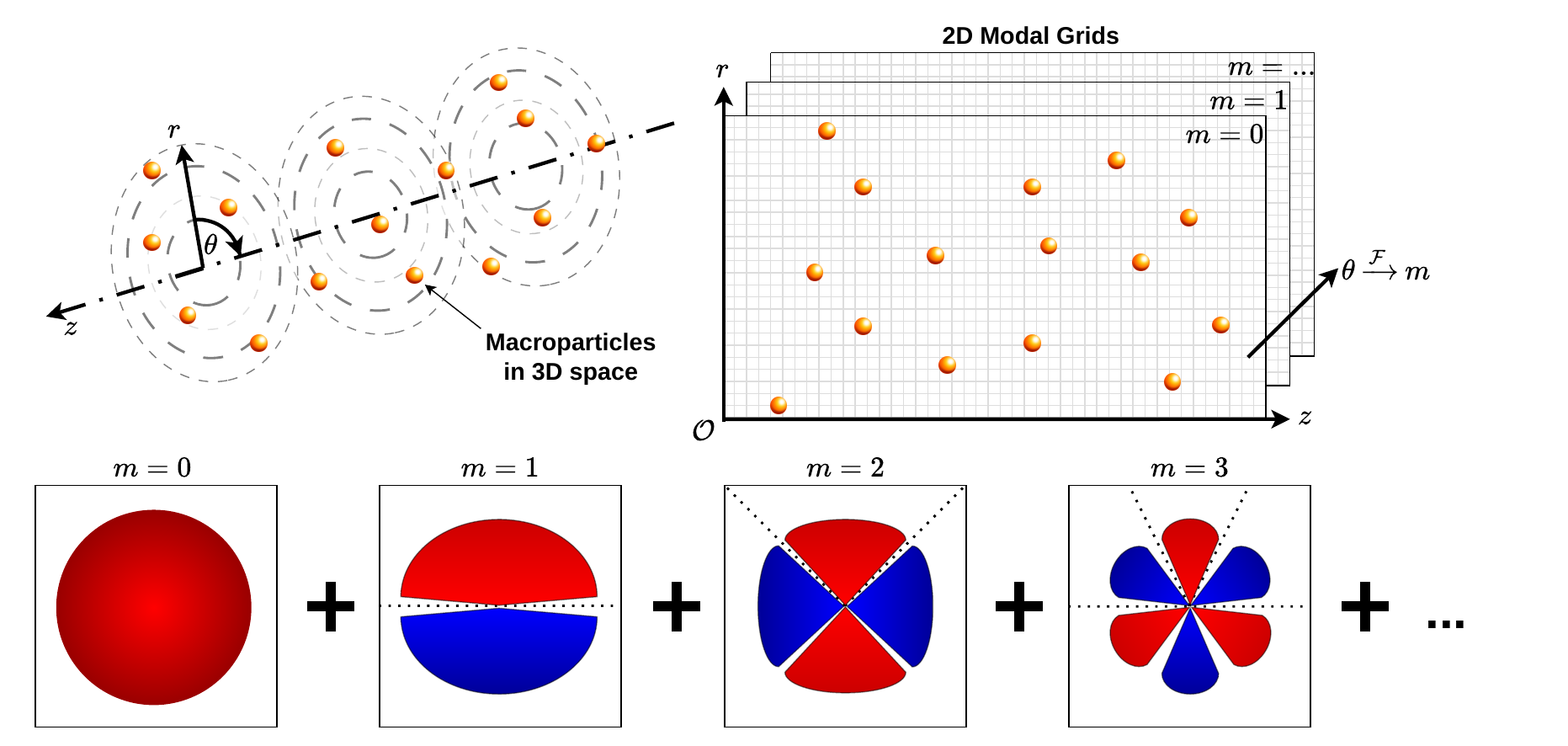}
\caption{Schematic of the DECOMPIC algorithm. All field quantities
are decomposed into $N_m+1$ complex azimuthal Fourier modes; each modal
amplitude satisfies an independent two-dimensional elliptic equation on the
meridional plane $\mathcal{D}$; and macroparticles are advanced in full
three-dimensional cylindrical coordinates with fields reconstructed by coherent
superposition of the modal contributions.}
\label{fig:decompic_schematic}
\end{figure*}

\subsection{Governing Equations and Physical Assumptions}
\label{sec:governing}

A nominally axisymmetric cylindrical plasma is considered with coordinate system $(r, \theta, z)$, where $r \in [0, r_{\max}]$ is the radial distance from the symmetry axis, $\theta \in [0, 2\pi)$ is the azimuthal angle and $z \in [0, L_z]$ is the axial coordinate. The physical domain is then the cylindrical volume $\Omega = \{(r,\theta,z) : 0 \leq r \leq r_{\max},\, \theta \in [0,2\pi),\, 0 \leq z \leq L_z\}$. Multiple particle species $s \in S$ are considered, each characterised by charge $q_s$, mass $m_s$, and a distribution function $f_s(\mathbf{x},\mathbf{v},t)$ defined on the six-dimensional particle phase space $(\mathbf{x}, \mathbf{v})$ parametrised by $(r, \theta, z, v_r, v_\theta, v_z)$. The evolution of each $f_s$ is therefore governed by the Vlasov equation
\begin{equation}
\frac{\partial f_s}{\partial t}
  + \mathbf{v}\cdot\nabla_{\mathbf{x}} f_s
  + \frac{q_s}{m_s}\bigl(\mathbf{E}+\mathbf{v}\times\mathbf{B}\bigr)
    \cdot\nabla_{\mathbf{v}} f_s = 0,
\label{eq:vlasov}
\end{equation}
coupled self-consistently to the fields through the charge density
$\rho = \sum_s q_s \int f_s\,d\mathbf{v}$ and current density
$\mathbf{j} = \sum_s q_s \int \mathbf{v} f_s\,d\mathbf{v}$.

The present implementation is electrostatic: the electric field derives from a scalar potential, $\mathbf{E} = -\nabla\phi$, with $\phi$ satisfying Poisson's equation $\nabla^2\phi = -\rho/\varepsilon_0$. This is appropriate when $\omega_p \ll c/L$, a condition met throughout the mesothermal, sub-relativistic regimes of interest in plasma propulsion and related low-temperature plasma devices. The magnetic field $\mathbf{B}$ is taken to be magnetostatic: a prescribed, time-independent external field with $\nabla\cdot\mathbf{B}=0$, with self-generated fields neglected on the grounds that the plasma beta is small.

\subsection{Azimuthal Fourier Decomposition}
\label{sec:fourier_decomp}

The fundamental structural ansatz of the formulation exploits the approximate rotational symmetry of the domain about the $z$-axis, so that departures of any sufficiently regular field from exact axisymmetry are representable by weighted complex Fourier modes in the azimuth. The three-dimensional domain $\Omega$ is fibred over the two-dimensional meridional plane $\mathcal{D} = (0,r_{\max}]\times[0,L_z]$ by the azimuthal circle $\theta \in [0,2\pi)$, and the natural function space for the field equations, $L^2(\Omega;\,r\,dr\,d\theta\,dz)$, admits the orthogonal decomposition

\begin{equation}
\begin{aligned}
&L^2(\Omega; r\,dr\,d\theta\,dz)
=
\bigoplus_{m\in\mathbb{Z}} \mathcal{H}_m,\\
\mathcal{H}_m
&=
\left\{
u(r,\theta,z)=\hat{u}_m(r,z)e^{-im\theta}
\;\middle|\;
\hat{u}_m\in L^2(\mathcal{D}; r\,dr\,dz)
\right\}.
\end{aligned}
\label{eq:hilbert_decomp}
\end{equation}

The three-dimensional field problem over $\Omega$ therefore becomes equivalent to the superposition of a family of two-dimensional $(r,z)$ problems over $\mathcal{D}$, one per modal subspace $\mathcal{H}_m$. The mode $m=0$ corresponds to the nominal (mean) axisymmetric component, while $m\neq0$ represents asymmetric or fluctuating components. 

Consistent with the decomposition of Eq.~(\ref{eq:hilbert_decomp}), the relevant field quantities are represented as truncated complex Fourier series in $\theta$:

\begin{align}
 & \phi(r, \theta, z, t) = \sum_{m \in \mathbb{Z}} \hat{\phi}_m(r, z, t)\, e^{-im\theta}, \label{eq:phi_series} \\
& \rho(r, \theta, z, t) = \sum_{m \in \mathbb{Z}} \hat{\rho}_m(r, z, t)\, e^{-im\theta}, \label{eq:rho_series}
\end{align}

where the complex modal amplitudes $\hat{\phi}_m,\hat{\rho}_m \in
L^2(\mathcal{D};\,r\,dr\,dz)$ are recovered by the $L^2([0,2\pi))$ projection
\begin{align}
  & \hat{\phi}_m(r,z,t)
  = \frac{1}{2\pi}\int_0^{2\pi} \phi(r,\theta,z,t)\, e^{im\theta}\, d\theta,
  \label{eq:phi_proj} \\
  & \hat{\rho}_m(r,z,t)
  = \frac{1}{2\pi}\int_0^{2\pi} \rho(r,\theta,z,t)\, e^{im\theta}\, d\theta.
  \label{eq:rho_proj}
\end{align}
Since $\phi$ and $\rho$ are real-valued, the conjugate symmetry $\hat{u}_{-m} = \hat{u}_m^*$ holds for all $m$, so only modes $m \geq 0$ need be retained explicitly, halving the storage and computational cost of the representation.

In numerical practice, truncating the infinite series at $ |m| \leq N_m $, for some finite $N_m \geq 0$, then retains $N_m+1$ independent copies of $\mathcal{D}$ with the truncation error controlled by the rate of spectral decay of $|\hat{\phi}_m|$ with $|m|$. For physical systems with smooth azimuthal structure --- as is typical of near-cylindrical plasma devices --- this decay is rapid and a small number of modes should suffice. It is worth emphasising at this point that the truncation to mode $N_m$ is a property of the field representation only: macroparticles are advanced in the full three-dimensional domain with fully self-consistent azimuthal positions $\theta_k \in [0, 2\pi)$, and their contribution to each mode is evaluated exactly via the projection in Eq.~(\ref{eq:rho_series}).

Before proceeding to the field equations, it is noted that the cylindrical field components $E_r$, $E_\theta$, $E_z$ carry the same azimuthal mode structure as $\phi$ itself. Substituting Eq.~(\ref{eq:phi_series}) into $\mathbf{E} =
-\nabla\phi$ and using $\partial_\theta e^{-im\theta} = -im\,e^{-im\theta}$
gives the modal field amplitudes
\begin{equation}
  \hat{E}_{r,m} = -\partial_r\hat{\phi}_m,
  \qquad
  \hat{E}_{\theta,m} = \frac{im}{r}\hat{\phi}_m,
  \qquad
  \hat{E}_{z,m} = -\partial_z\hat{\phi}_m.
  \label{eq:Efields}
\end{equation}
The azimuthal component acquires the factor $im/r$ --- the Fourier-space
representation of the azimuthal gradient operator $r^{-1}\partial_\theta$ ---
which is exact at every mode order and incurs no additional discretisation
error.

\subsection{Modal Decomposition of the Poisson Equation}
\label{sec:modal_poisson}
The Poisson equation $\nabla^2\phi = -\rho/\varepsilon_0$, expressed in cylindrical coordinates, takes the form

\begin{equation}
\frac{1}{r}\frac{\partial}{\partial r}\!\left(r\frac{\partial \phi}{\partial r}\right) + \frac{1}{r^2}\frac{\partial^2 \phi}{\partial \theta^2} + \frac{\partial^2 \phi}{\partial z^2} = -\frac{\rho}{\varepsilon_0}.
\label{eq:poisson}
\end{equation}

Substituting the representations Eqs.~(\ref{eq:phi_series}) and (\ref{eq:rho_series}) into Eq.~(\ref{eq:poisson}) gives

\begin{align}
\sum_{m \in \mathbb{Z}} \Bigg[\frac{1}{r}\frac{\partial}{\partial r}\!\left(r\frac{\partial \hat{\phi}_m}{\partial r}\right)e^{-im\theta} + \frac{1}{r^2}\frac{\partial^2}{\partial \theta^2}\left(\hat{\phi}_m e^{-im\theta}\right) \label{eq:poissonexpanded}\\ \notag  + \frac{\partial^2 \hat{\phi}_m}{\partial z^2}e^{-im\theta}\Bigg] = -\frac{1}{\varepsilon_0}\sum_{m \in \mathbb{Z}} \hat{\rho}_m\, e^{-im\theta}.
\end{align}

The azimuthal second derivative acting on the $m$-th term yields $\partial^2_\theta(\hat{\phi}_m e^{-im\theta}) = -m^2 \hat{\phi}_m e^{-im\theta}$, since $\hat{\phi}_m$ is independent of $\theta$ by construction. {\color{black}Eq.~(\ref{eq:poissonexpanded})} therefore becomes

\begin{align}
\sum_{m \in \mathbb{Z}} \Bigg[\frac{1}{r}\frac{\partial}{\partial r}\!\left(r\frac{\partial \hat{\phi}_m}{\partial r}\right) - \frac{m^2}{r^2}\hat{\phi}_m + \frac{\partial^2 \hat{\phi}_m}{\partial z^2}\Bigg] e^{-i m\theta}\\ \notag = -\frac{1}{\varepsilon_0}\sum_{m \in \mathbb{Z}} \hat{\rho}_m\, e^{-i m\theta}.
\end{align}

To isolate the equation for a single mode, both sides are multiplied by $e^{im'\theta}$ for a fixed projection index $m' \neq m$ and integrated over $\theta \in [0, 2\pi)$ to project onto the $m'$-th mode. Using the orthogonality relation $\frac{1}{2\pi}\int_0^{2\pi} e^{i(m'-m)\theta}d\theta = \delta_{m'm}$, all cross terms vanish identically. This eliminates every term in the sum except $m=m'$, after which the prime is dropped, yielding independent two-dimensional elliptic equations for each azimuthal mode $m$:

\begin{equation}
\frac{1}{r}\frac{\partial}{\partial r}\!\left(r\frac{\partial \hat{\phi}_m}{\partial r}\right) - \frac{m^2}{r^2}\hat{\phi}_m + \frac{\partial^2 \hat{\phi}_m}{\partial z^2} = -\frac{\hat{\rho}_m}{\varepsilon_0},
\label{eq:modal_poisson}
\end{equation}

which may be written compactly as $\mathcal{L}_m\hat\phi_m = -\hat\rho_m/\varepsilon_0$,
where
\begin{equation}
  \mathcal{L}_m
  \equiv \mathcal{B}_m + \partial_z^2
  = \underbrace{\frac{1}{r}\frac{\partial}{\partial r}\!\left(r\frac{\partial}{\partial r}\right)
    - \frac{m^2}{r^2}}_{\text{Bessel operator }\mathcal{B}_m}
  + \;\frac{\partial^2}{\partial z^2},
  \label{eq:Lm_def}
\end{equation}

This is the central result of the theoretical formulation. The three-dimensional Poisson problem over $\Omega$ is reduced exactly to a family of $N_m + 1$ independent two-dimensional elliptic problems on the meridional plane $\mathcal{D}$, one per subspace $\mathcal{H}_m$, each driven by a single complex modal source $\hat\rho_m$. The decoupling is an exact consequence of the Fourier orthogonality and imposes no restriction on the structure of $\hat\rho_m$; it holds for any $\rho$ admitting the representation of Eq.~(\ref{eq:rho_series}), regardless of mode amplitudes. The scheme reduces to a standard axisymmetric ($r,z$) PIC code in the limit $N_m = 0$.

The operator $\mathcal{L}_m$ in Eq.~(\ref{eq:Lm_def}) has a transparent structure: its radial part $\mathcal{B}_m \equiv r^{-1}\partial_r(r\partial_r) - m^2/r^2$ is the Bessel operator of order $m$, which differs from the axisymmetric ($m=0$) cylindrical Laplacian by the centrifugal shift $-m^2/r^2$, and is augmented in the $z$-direction by the standard second derivative $\partial_z^2$. Each discrete $m^2/r^2$ term contributes a non-negative diagonal increment to every cell of the corresponding assembled operator matrix $\mathsf{L}_m$, that grows with $m$, which has two important consequences relative to the $m=0$ case. 

First, the Weyl monotonicity bound applied to the discrete splitting $\mathsf{L}_m =
\mathsf{L}_0 - \mathrm{diag}(m^2/r^2)$ gives
\begin{equation}
  \kappa(\mathsf{L}_m)
  \leq \frac{|\lambda_{\max}(\mathsf{L}_0)| + m^2/\Delta r_{\min}^2}{
             |\lambda_{\min}(\mathsf{L}_0)| + m^2/r_{\max}^2}
  \;\xrightarrow{\;m\to\infty\;}\; \frac{r_{\max}^2}{\Delta r_{\min}^2},
  \label{eq:cond_number}
\end{equation}
where $\lambda$ represents an eigenvalue. The matrix condition number $\kappa$ therefore decreases monotonically with $m$ and saturates at a value independent of the axial domain length. The conditioning analysis suggests progressively improved conditioning for higher mode numbers.
 
Secondly, the behaviour of $\hat\phi_m$ near the axis is determined by the indicial analysis of Eq.~(\ref{eq:modal_poisson}) as $r \to 0$, which gives $\hat\phi_m \sim r^{|m|}$, enforcing the axis conditions
\begin{equation}
  \left.\partial_r\hat\phi_0\right|_{r=0} = 0
  \;\;(m=0),
  \qquad
  \hat\phi_m\big|_{r=0} = 0
  \;\;(m \geq 1).
  \label{eq:axis_bc}
\end{equation}
These express the physical requirement that a mode of azimuthal order $m \geq 1$ must carry zero amplitude on the axis --- a non-zero value would imply an infinite azimuthal gradient --- and is the discrete counterpart of the regularity condition required by the Bessel functions. 

Up to this point, it is noted the decomposition is exact. The only approximation introduced by the present method is truncation of the Fourier expansion at a finite maximum mode number $N_m$. Particle trajectories remain fully three-dimensional and all retained modes contribute self-consistently to the reconstructed electric field experienced by each particle.

\subsection{Charge Density Decomposition and Particle Deposition}
\label{sec:deposition}
{\color{black}For macroparticles with weight $W$ (all macroparticles are equally weighted) and thus macro-charge $Q_s = W q_s$}, the modal charge density at meridional grid node $(r_p,z_q)$ is obtained by projecting the particle ensemble onto mode $m$ via {\color{black}Eq.~(\ref{eq:rho_proj})}:
\begin{equation}
  \hat{\rho}_m(r_p,z_q,t)
  = \frac{1}{V_{p,q}} \sum_s\sum_{k\in\mathcal{N}(p,q)}
    Q_s\,\mathcal{S}(r_p{-}r_k, z_q{-}z_k)\,e^{im\theta_k},
  \label{eq:rho_deposition}
\end{equation}
where $V_{p,q}$ is the volume of the cylindrical annular cell at node
$(p,q)$, and $\mathcal{N}(p,q)$ denotes the set of macroparticles within
the compact support of $\mathcal{S}$ centred on $(r_p,z_q)$. Each such macroparticle
contributes simultaneously to all retained modes through the complex
phase factor $e^{im\theta_k}$, the only quantity in
Eq.~(\ref{eq:rho_deposition}) that encodes the particle's azimuthal
position. The shape function weights $\mathcal{S}(r_p{-}r_k,z_q{-}z_k)$
are computed once per particle and reused across all modes, so the
marginal cost per additional mode is a single complex multiply.

\begin{figure*}[!t]
\centering
\includegraphics[width=\textwidth]{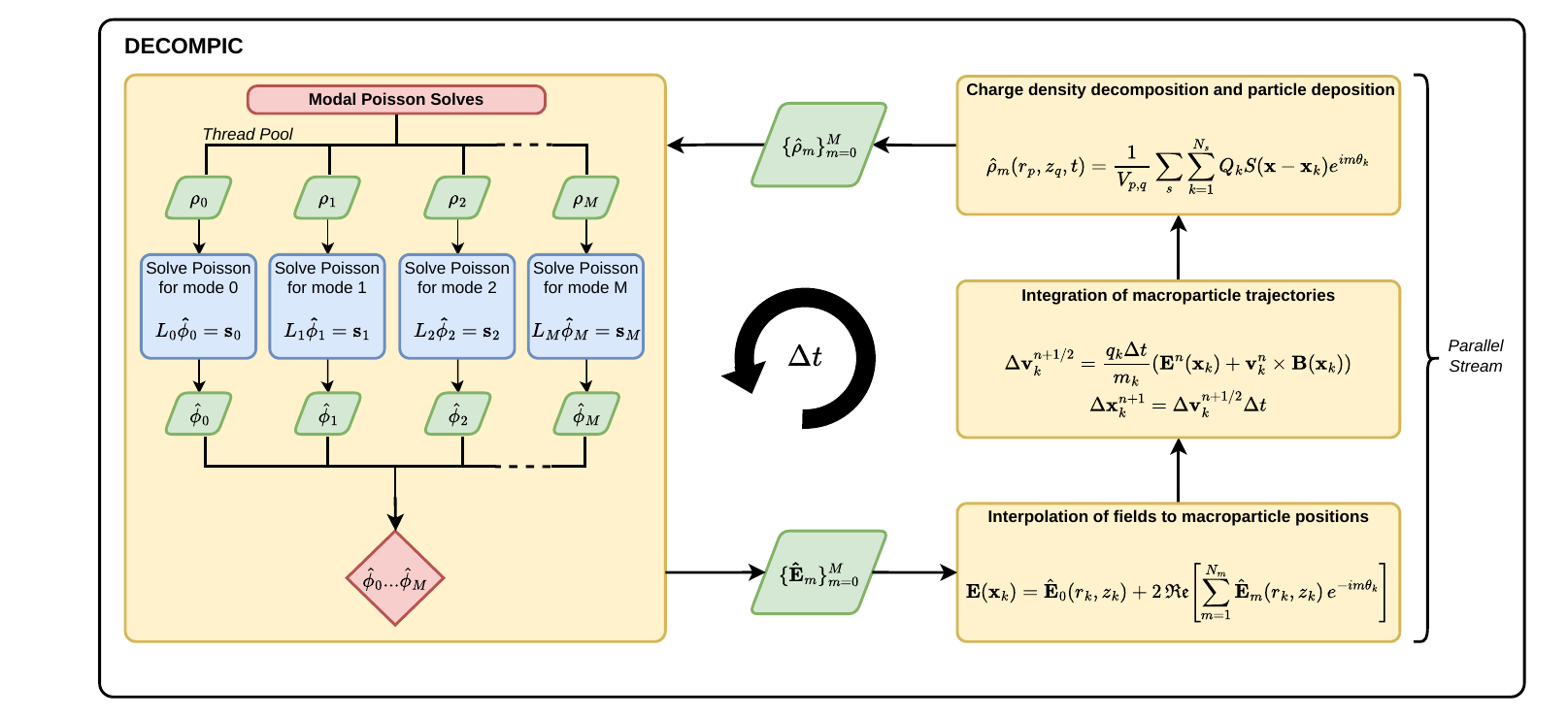}
\caption{Structure of a single DECOMPIC time-step. Charge deposition projects
the three-dimensional particle ensemble onto the $N_m + 1$ independent
meridional grids; the modal Poisson solves run concurrently across a thread
pool; the reconstructed three-dimensional fields drive the Boris particle
advance.}
\label{fig:pic_cycle}
\end{figure*}

The phase factor is evaluated without computing $\theta_k$ explicitly.
Writing the Cartesian particle position as
$x_k + iy_k = r_k e^{i\theta_k}$,
\begin{equation}
  e^{im\theta_k}
  = \left(\frac{x_k + iy_k}{r_k}\right)^m,
  \label{eq:phase}
\end{equation}
which avoids the branch-cut ambiguity of $\arctan(y_k/x_k)$ near the
axis and recovers unity for $m=0$, giving the standard axisymmetric
deposition. The sequence $\{e^{im\theta_k}\}_{m=0}^{N_m}$ is generated
by the angle-addition recurrence
\begin{equation}
  e^{i(m+1)\theta_k} = e^{im\theta_k}\cdot e^{i\theta_k},
  \label{eq:phase_recurrence}
\end{equation}
requiring one complex multiply per mode with no further transcendental
function evaluations.

\subsection{Field Reconstruction and Particle Interpolation}
\label{sec:interpolation}
Given the modal potentials $\{\hat{\phi}_m\}_{m=0}^{N_m}$ obtained by solving Eq.~(\ref{eq:modal_poisson}), the three-dimensional electric field on $\Omega$ at
any point $(r_p,\theta,z_q)$ is reconstructed from Eq.~(\ref{eq:Efields}) as
\begin{equation}
  \mathbf{E}(r_p,\theta,z_q,t)
  = \hat{\mathbf{E}}_0(r_p,z_q)
    + 2\,\mathfrak{Re}\!\left[\sum_{m=1}^{N_m}
        \hat{\mathbf{E}}_m(r_p,z_q)\,e^{-im\theta}\right],
  \label{eq:field_reconstruct}
\end{equation}
where the factor of 2 and the real-part operation account for the
conjugate symmetry of the negative modes. The electric field at
macroparticle $k$ is instead obtained by first interpolating each modal
amplitude from the surrounding grid nodes to $(r_k,z_k)$ using the
same shape function as the deposition step, then applying the azimuthal
phase. {\color{black}Using the same $\mathcal{S}$ as in Eq.~(\ref{eq:rho_deposition}) ensures that the interpolation and deposition operators are adjoints, providing a consistent particle-field coupling. This avoids systematic self-forces associated with mismatched deposition and interpolation weights, which is important for suppressing artificial numerical heating and cooling~\cite{birdsall1991}:}
\begin{equation}
\begin{aligned}
  \mathbf{E}(\mathbf{x}_k)
  &= \sum_{(p,q)\in\mathcal{G}(k)}
       \mathcal{S}(r_p{-}r_k, z_q{-}z_k)\,
       \hat{\mathbf{E}}_0(r_p,z_q) \\
  &\quad +\, 2\,\mathfrak{Re}\!\Biggl[\sum_{m=1}^{N_m}
      \biggl(\sum_{(p,q)\in\mathcal{G}(k)}
        \mathcal{S}(r_p{-}r_k, z_q{-}z_k) \\
  &\qquad\qquad\qquad\qquad\quad\times\,
        \hat{\mathbf{E}}_m(r_p,z_q)
      \biggr) e^{-im\theta_k}
    \Biggr],
\end{aligned}
  \label{eq:field_interp}
\end{equation}
where $\mathcal{G}(k)$ denotes the stencil of grid nodes in the meridional plane within the support of $\mathcal{S}$ surrounding particle $k$. The phase $e^{-im\theta_k}$ is accumulated using the same recurrence Eq.~(\ref{eq:phase_recurrence}) as the deposition step, adding no further transcendental function evaluations.

The cylindrical field components are converted to Cartesian form via
$E_x = E_r\cos\theta_k - E_\theta\sin\theta_k$,
$E_y = E_r\sin\theta_k + E_\theta\cos\theta_k$, $E_z = E_z$,
and the particle trajectories are then integrated using the standard Boris algorithm.

The scheme is fully self-consistent: modal sources $\{\hat\rho_m\}$ are deposited from the three-dimensional particle ensemble, the $N_m+1$ independent systems of Eq.~(\ref{eq:modal_poisson}) are solved, and the reconstructed fields from Eq.~(\ref{eq:field_interp}) drive the particle advance. Global charge conservation follows from the linearity of the deposition: $\sum_k Q_k e^{im\theta_k}$ for $m \neq 0$ contributes nothing to the volume-integrated charge by orthogonality, while $m=0$ recovers the exact total charge exactly.



\section{Numerical Implementation}
\label{sec:implementation}

\subsection{Code Architecture and Parallelisation Strategy}
\label{sec:parallelisation}

The method is implemented in a parallelised Java codebase, hereafter referred to as DECOMPIC. Two levels of concurrency arise naturally from the mathematical structure of Section~\ref{sec:theory}. At the outermost level, the $N_m+1$ modal Poisson solves are entirely independent: mode $m$ takes only $\hat{\rho}_m$ as input and produces $\hat{\phi}_m$ as output, with no data dependencies between subspaces $\mathcal{H}_m$ at the field-solve stage. These solves are therefore dispatched as independent tasks across a thread pool. Load-balancing is maintained by assigning modal pairs to threads symmetrically around the midpoint of the mode index range, exploiting the property that convergence cost should favourably decrease with $m$ per the conditioning bound established in Eq.~(\ref{eq:cond_number}). For $N_m+1$ modes distributed across $N_T$ threads, the optimal assignment groups mode $m$ with mode $N_m-m$ on the same thread, so that thread $t \in \{0,\ldots,N_T-1\}$ is assigned the mode pair $\{t,\, N_m - t\}$. More generally, with $N_T \leq \lfloor(N_m+2)/2\rfloor$, thread $t$ is assigned the interleaved mode set

\begin{align}  
\mathcal{M}_t = \{t + kN_T \;:\; k \in \mathbb{Z}_{\geq 0}, \; t + kN_T \leq N_m\} \\ \nonumber \cup \{N_m - t - kN_T \;:\;  k \in \mathbb{Z}_{\geq 0}, \; N_m - t - kN_T \geq 0\},
\end{align}

distributing the most expensive low-$m$ solves evenly with the cheapest high-$m$ solves. The expected wall-clock time per thread is then approximately equalised across the pool because the improved conditioning of high modes compensates for the full cost of the low modes they are grouped with. At the inner level, the particle advance, charge deposition, and field interpolation kernels are parallelised over the macroparticle ensemble using parallel stream abstraction. The full PIC cycle is summarised in Figure~\ref{fig:pic_cycle}.
\subsection{Stability and Numerical Constraints}
\label{sec:constraints}
 
The resolution and stability requirements on the spatial grid, time-step, and particle loading all follow from the structure of the decomposition and are derived here together.
 
It is first noted that the $m \geq 1$ modal solves are driven by the perturbation component of the density. The normalised amplitude $\tilde{\epsilon}_{s,m} = |\hat{n}_{s,m}|/\hat{n}_{s,0}$, where $\hat{n}_{s,0}(r,z)$ is the local azimuthal mean density, satisfies $\tilde{\epsilon}_{s,m} \in [0,1]$ rigorously: the triangle inequality applied to the Fourier projection~(\ref{eq:phi_proj}) with $|e^{im\theta}|=1$ gives $|\hat{n}_{s,m}| \leq \hat{n}_{s,0}$ for all $m$. The bound is approached but not exceeded even in strongly non-uniform configurations.
 
\subsubsection{Time-step constraints}
\label{sec:timestep}
 
The standard PIC stability requirements on electron plasma frequency, Courant condition, and gyro-frequency give the operational time-step
\begin{equation}
    \Delta t \leq \min\!\left(
        \frac{\color{black}K}{\omega_{pe}^{(0)}},\;
        \frac{\Delta r}{u_e + 3v_{th,e}},\;
        \frac{\Delta z}{u_e + 3v_{th,e}},\;
        \frac{0.35}{\Omega_{ce}}
    \right),
    \label{eq:dt_constraint}
\end{equation}
where $v_{th,e} = \sqrt{k_BT_e/m_e}$ is the electron thermal speed, {\color{black}$u_e$ is the electron drift speed}, and $\Omega_{ce} = eB/m_e$ is the electron cyclotron frequency. {\color{black}For the plasma-frequency constraint, $K$ is a stability coefficient that sets the allowable fraction of the plasma oscillation period resolved by the time-step. A conservative value often adopted in explicit electrostatic PIC is $K\simeq0.1$, although stable operation with $K$ in the range {\color{black}0.05--0.3} is commonly reported depending on the numerical formulation and the plasma regime{\color{black}\cite{birdsall1991,picreview}}.

 Since the modal electron density coefficients $\hat{n}_{e,m}$ represent perturbations of the same underlying equilibrium plasma, rather than independent equilibrium plasma populations, the plasma frequency constraint is common to all Fourier modes. Consequently, linearising the plasma response about the equilibrium density $\hat{n}_{e,0}$ gives the same characteristic plasma frequency scale for each azimuthal Fourier component: the modal amplitude determines the strength of the response, while the restoring force is set by the equilibrium density. The detailed linearised derivation is given in Appendix~\ref{sec:appendixA}. Thus, Eq.~(\ref{eq:dt_constraint}) governs all modes; the Fourier decomposition does not introduce a separate time-step constraint for each mode.}
 
An additional azimuthal Courant condition arises from the finite bandwidth of the mode representation. The highest retained mode $N_m$ can represent azimuthal structure only down to a wavelength $2\pi r / N_m$ at radius $r$; a particle traversing more than one half-wavelength per step aliases energy into unresolved modes. This defines an effective azimuthal cell size
\begin{equation}
    \Delta\theta^{\rm eff}(r) = \frac{\pi r}{N_m},
    \label{eq:virtual_cell}
\end{equation}
and the associated Courant streaming constraint
\begin{equation}
    \Delta t \leq \frac{\pi\, r}{N_m\,v_{\theta,\max}},
    \label{eq:azimuthal_cfl}
\end{equation}
This is generally less restrictive than Eq.~(\ref{eq:dt_constraint}) at typical device radii, but should be checked when large azimuthal drift velocities are present. The full operational time-step is thus the minimum of Eq.~(\ref{eq:dt_constraint})
and Eq.~(\ref{eq:azimuthal_cfl}).
 
\subsubsection{Spatial constraints and radial mesh requirements}
\label{sec:grid}
 
{\color{black}The Debye constraint $\Delta r, \Delta z \lesssim \lambda_{D,0}$ for the $m=0$ background automatically satisfies grid-heating stability for all higher modes, since physical Debye length is determined by the equilibrium plasma density,
$
\lambda_{D,0}
=
\sqrt{{\epsilon_0 k_BT_e}/{\hat{n}_{e,0}e^2}}.
$ As with the plasma frequency, the Debye length is a property of the equilibrium plasma and is common to the perturbation modes in the linearised response, as shown in Appendix \ref{sec:appendixA}. Two further resolution requirements are specific to the Fourier decomposition however and bear directly on the design of the mesh.}
 
The virtual cell of Eq.~(\ref{eq:virtual_cell}) imposes a radial consistency condition: for the meridional grid not to be coarser than the azimuthal structure to resolve at the same radius, one requires
\begin{equation}
  \frac{\Delta r_p}{r_p} \lesssim \frac{\pi}{N_m}.
  \label{eq:radial_consistency}
\end{equation}
This is a fractional constraint on the cell size, demanding geometric spacing throughout the domain and therefore violated near the axis by any uniform mesh with fixed $\Delta r$.
 
A more demanding requirement arises from the structure of $\mathcal{L}_m$. The modal eigenfunctions near the axis behave as $\hat\phi_m \sim r^{|m|}$, so the radial gradient scales as $\partial_r\hat\phi_m \sim |m|r^{|m|-1}$. For a finite-volume stencil to represent this profile to second-order accuracy, the truncation error in the centred-difference approximation to $\partial_r\hat\phi_m$ scales as $\Delta r^2\,\partial_{rrr}\hat\phi_m \sim \Delta r^2\,|m|(|m|-1)(|m|-2)\,r^{|m|-3}$. Requiring this to be small relative to the gradient itself gives $\Delta r^2/r^2 \lesssim 1/(|m|^2)$ and the fractional constraint
\begin{equation}
  \frac{\Delta r_p}{r_p} \lesssim \frac{1}{N_m}.
  \label{eq:eigenfunction_constraint}
\end{equation}
This is stricter than Eq.~(\ref{eq:radial_consistency}) by a factor of $\pi$, but applies only near the axis. For $m \geq 1$ the Dirichlet condition $\hat\phi_m|_{r=0} = 0$ is imposed exactly via a boundary row substitution, thus the accuracy argument is not required to hold at $p=0$ itself. The binding constraint is at the first interior cell $p=1$ and is again violated by a uniform mesh. By contrast, condition Eq.~(\ref{eq:radial_consistency}) applies to the full mesh and sets a weaker but globally uniform requirement. Together they motivate a radial mesh that clusters cells near the axis, with the required degree of clustering scaling directly with $N_m$.

\subsubsection{Signal-to-noise ratio and particle-per-cell requirements}
\label{sec:snr}
 
Because all modal subspaces $\mathcal{H}_m$ share the same macroparticle
ensemble, the noise has a single origin:
the finite-sample representation of the continuous distribution function.
For a cell containing $N_{\rm ppc}$ macroparticles with azimuthal positions
$\{\theta_k\}$ drawn from the physical distribution, the deposited amplitude
of mode $m$ is estimated by the sample mean $A_m =
N_{\rm ppc}^{-1}\sum_k e^{im\theta_k}$. The {\color{black}species-density} perturbation at
mode $m$ has normalised amplitude $\tilde\epsilon_{s,m} = |\hat n_m|/\hat n_0 \in
[0,1]$, which is the signal: $\langle e^{im\theta}\rangle_g =
\tilde\epsilon_{s,m} e^{i\varphi_m}$. Since $|e^{im\theta}|=1$ identically, the
variance of a single sample is $\sigma_{s,m}^2 = 1 - \tilde\epsilon_{s,m}^2$,
giving the per-mode {\color{black}species-density} signal-to-noise ratio
\begin{equation}
    \mathrm{SNR}_{{\color{black}s},m}
    = \tilde\epsilon_{s,m}\sqrt{\frac{N_{\rm ppc}}{1-\tilde\epsilon_{s,m}^2}},
    \label{eq:snr}
\end{equation}
which for $\tilde\epsilon_{s,m} \ll 1$ reduces to $\tilde\epsilon_{s,m}
\sqrt{N_{\rm ppc}}$, consistent with Lifschitz et al.~\cite{lifschitz2009}. The requirement
$\mathrm{SNR}_{{\color{black}s},m} \geq \mathrm{SNR}_{\min}$ gives
\begin{equation}
    N_{\rm ppc} \geq
    \frac{\mathrm{SNR}_{\min}^2\,(1 - \tilde\epsilon_{s,m}^2)}{\tilde\epsilon_{s,m}^2}.
    \label{eq:nppc_requirement}
\end{equation}
For a growing instability, this requirement need only be satisfied at the
amplitude of physical interest rather than at the seeded perturbation level:
an exponentially growing mode with initial amplitude $ \ll 1$ and
growth rate $\gamma$ rises above the noise floor $1/\sqrt{N_{\rm ppc}}$
within a few growth times regardless of $N_{\rm ppc}$, becoming self-resolving
before it reaches saturation. Eq.~(\ref{eq:nppc_requirement}) is
therefore binding primarily for steady-state or weakly driven modes that must
be resolved at a fixed amplitude. At saturation, where $\tilde\epsilon_{s,m}
\sim \mathcal{O}(0.01$--$1)$, the requirement is modest: for $\tilde\epsilon_{s,m}
= 0.3$ and $\mathrm{SNR}_{\min} = 3$ one needs $N_{\rm ppc} \geq 91$. This
particle cost is not specific to the DECOMPIC method: any simulation resolving
the same fractional perturbation faces an identical shot-noise constraint.

{\color{black}The quantity entering Poisson's equation is instead the net charge density, $\hat{\rho}_m=\sum q_s\hat{n}_{s,m}$. Its statistical noise is determined by the charge-weighted contributions from the different particle species and is distinct from the SNR$_{s,m}$ of an individual species-density mode. For statistically independent particle populations of comparable loading at $N_{ppc}$, the corresponding noise variance is $\sigma^2_{\rho,m}=\sum_s (q_s n_{s,0})^2 \left( 1-\tilde{\epsilon}_{s,m}^2 \right){ /N_{ppc} }$. Since the equilibrium net charge density, $\hat{\rho}_0$, may be close to zero in a quasi-neutral plasma $\hat{n}_{i,0}\simeq \hat{n}_{e,0}=n_0$, the charge-density perturbation is instead normalised to the characteristic single-species charge density, $\tilde{\epsilon}_{\rho,m}=|\hat{\rho}_m|/(|q_e|n_0)$. In the small-perturbation limit, $\sigma_{\rho,m}\approx |q_e| n_0 \sqrt{{2}/{N_{\rm ppc}}}$ and the corresponding SNR is then

\begin{equation}
{\rm SNR}_{\rho,m}
\approx
\tilde{\epsilon}_{\rho,m}
\sqrt{
\frac{N_{\rm ppc}}{2}
}.
\label{eq:snrrho}
\end{equation}

Hence, the species-density and charge-density SNRs quantify two distinct but related resolutions: Eq. (\ref{eq:snr}) measures the statistical resolution of the underlying individual species-density modes, while Eq. (\ref{eq:snrrho}) quantifies the statistical resolution of the net charge density modal source that directly drives Poisson's equation. The effect of the latter on the electric field result, however, is determined by the modal Poisson operator which acts as a mode-dependent spatial filter on the charge-density noise. In particular, the azimuthal contribution $-\frac{m^2}{r^2}\hat{\phi}_m$ increasingly suppresses the potential response to charge-density noise fluctuations at higher azimuthal mode numbers. If considering an azimuthally dominated limit, the corresponding responses scale approximately as $\hat{\phi}_m \propto {\hat{\rho}_m}/{m^2}, \quad \hat{E}_{m} \propto {\hat{\rho}_m}/{m}$. The precise charge-to-field noise transfer, however, depends on the full radial structure, boundary conditions, discretisation, and spatial distribution of the charge-density fluctuations, and therefore cannot be characterised by a single general SNR estimate here. Nevertheless, it follows that while the noise in $\hat{\rho}_m$ does not necessarily decrease with $m$,  the response of the potential and electric field to a given charge-density noise amplitude does. The modal Poisson solve therefore does not introduce any stringent $m$-dependent $N_{\mathrm{ppc}}$ requirement for the charge-to-field noise. Consequently, the charge-density $\mathrm{SNR}_{\rho,m}$ is not a significant additional constraint on the particle loading and the underlying solution noise remains bound by standard PIC $\sqrt{N_{ppc}}^{-1}$ scaling.}

{\color{black}For applications requiring accurate reconstruction of the full azimuthal parameter profiles, the relevant quantity is the noise in the reconstructed fields $u(r_p,\theta,z_q)$. In the small-perturbation regime, where the fluctuating quantity $\delta u/u_0$ remains sufficiently small, the noise contributions from different modes are uncorrelated to leading order in $\tilde{\epsilon}_{u,m}=|\hat{u}_m|/\hat{u}_0$. In this limit, the cross-covariance $\mathrm{Cov}(\delta\hat{u}_m,\delta\hat{u}_{m'}^*)$ is $\mathcal{O}(\tilde{\epsilon}_{u,m}^2/N_{\mathrm{ppc}})$ for $m\neq m'$, and each of the $1+2N_m$ independent real Fourier components contributes an independent noise term to the reconstruction. The mean-squared noise averaged over $\theta$ is therefore
\begin{equation}
    \langle|\delta u|^2\rangle_\theta
    \approx \frac{1 + 2N_m}{N_{\rm ppc}},
    \label{eq:recon_noise}
\end{equation}
and the reconstruction SNR is
\begin{equation}
    \mathrm{SNR}_{\rm rec}
    \approx \left(\frac{N_{\rm ppc}}{1 + 2N_m}\right)^{1/2}.
    \label{eq:snr_rec}
\end{equation}
This scaling applies generally to any variable $u \in \left\{ \rho, n_s, \phi, \mathbf{E}, \ldots \right\},$ represented by the azimuthal Fourier decomposition. Maintaining a fixed reconstruction noise level as modes are added requires $N_{\rm ppc} \propto 1 + 2N_m$, which is the Nyquist criterion for the reconstruction: $2N_m$ independent azimuthal samples are the minimum needed
to distinguish $N_m$ harmonics, consistent with the empirical finding of
Lehe~\textit{et al.}~\cite{lehe2016}. For $N_m = 4$ and $\mathrm{SNR}_{\rm
rec} = 10$ this requires only $N_{\rm ppc} \approx 900$, a standard loading
for explicit PIC. }

{\color{black}It is important to highlight however that this approximation may not hold for strongly excited large-scale modes. In the presence of large-amplitude azimuthal structures, such as a rotating spoke, the fluctuating modal components of the charge density $\hat{\rho}_m$ can become sufficiently large that cross-correlations between Fourier modes are no longer negligible. In this regime, Eq. (\ref{eq:recon_noise}) should therefore be regarded as a limited approximation whose accuracy must be assessed for the specific discharge conditions, rather than as a generally valid expression for the full reconstructed-field noise.}

The more demanding per-mode requirements of Eq.~(\ref{eq:nppc_requirement})
govern when $\tilde\epsilon_{s,m} < 1/\sqrt{1+2N_m}$; otherwise Eq.~(\ref{eq:snr_rec})
is binding.

{\color{black}Separately from shot noise, finite-$N_{ppc}$ loading is also subject to numerical thermalisation, a distinct artefact that can cause particle velocity distributions to evolve toward a Maxwellian as macroparticles experience polarisation drag and resonantly interact with the fluctuation spectrum\cite{jubin}. This effect is not addressed quantitatively here; the particle counts used in Sec. \ref{sec:benchmarking} fall within the range where such effects have been reported and should be kept in mind alongside the requirements above when selecting $N_{ppc}$, especially for cases over many thousands of plasma periods.}

\subsection{Finite-Volume Discretisation on a Stretched Radial Mesh}
\label{sec:fv}
 
{\color{black}The radial grid is defined using a cell-centred finite-volume
discretisation. The cell faces (located at grid points $r_{p+1/2}$) are obtained from the
hyperbolic-tangent mapping
\begin{align}
r_{p+1/2}
=
r_{\max}
\frac{
\tanh\!\left[\beta(\xi_{p+1/2}-\xi_c)\right]
-
\tanh(-\beta\xi_c)
}{
\tanh\!\left[\beta(1-\xi_c)\right]
-
\tanh(-\beta\xi_c)
},
\\ 
\xi_{p+1/2}=\frac{p+1/2}{N_r}, \notag
\end{align}
where $\beta > 0$ controls the degree of axis clustering and $\xi_c \in
(0,1)$ locates the target clustering region. The stretching parameter is
chosen so that adjacent cell-size ratios do not exceed $1.2$--$1.5$,
maintaining second-order accuracy of the stencil and ensuring the
eigenfunction of Eq.~(\ref{eq:eigenfunction_constraint}) and virtual-cell
of Eq. (\ref{eq:radial_consistency}) conditions are met. Cell centres (located at grid points $r_p$) and widths are
\begin{align}
r_p = \tfrac{1}{2}(r_{p-1/2}+r_{p+1/2}), \qquad
\Delta r_p = r_{p+1/2}-r_{p-1/2},
    \label{eq:grid_def}
\end{align}
with inter-cell-centre distances $d_{p\pm 1/2} = r_{p\pm 1} - r_p$ unequal in general on the stretched grid. The axial direction uses a uniform mesh with spacing $\Delta z = L_z / N_z$.
 
The modal Poisson equation of Eq.~(\ref{eq:modal_poisson}) is discretised by a cell-centred finite-volume method in its $r$-weighted conservative form, $r\mathcal{L}_m$. Integrating this form over the control volume $\mathcal{V}_{p,q} = [r_{p-1/2},r_{p+1/2}]\times[z_{q-1/2},z_{q+1/2}]$, the radial divergence term is evaluated from the face fluxes, with face-centred derivatives approximated by second-order centred differences using the inter-cell-centre distances $d_{p\pm 1/2}$. The axial and centrifugal volume integrals are evaluated using cell-centre quadrature. This yields the five-point stencil
\begin{align}
    \alpha_{p+1}^{(m)}\hat{\phi}_{m,p+1,q}
    &+ \alpha_{p-1}^{(m)}\hat{\phi}_{m,p-1,q}
    + \beta_p\!\left(\hat{\phi}_{m,p,q+1} + \hat{\phi}_{m,p,q-1}\right)
    \nonumber \\
    &- A_{p,q}^{(m)}\,\hat{\phi}_{m,p,q}
    = -\frac{r_p\,\hat{\rho}_{m,p,q}}{\varepsilon_0},
    \label{eq:stencil}
\end{align}
with coefficients
\begin{align}
    \alpha_{p\pm 1}^{(m)} &= \frac{r_{p\pm 1/2}}{\Delta r_p\,d_{p\pm 1/2}},
    \qquad
    \beta_p = \frac{r_p}{\Delta z^2},
    \label{eq:off_diagonal} \\[4pt]
    A_{p,q}^{(m)} &= \alpha_{p+1}^{(m)} + \alpha_{p-1}^{(m)} + 2\beta_p
    + \underbrace{\frac{m^2}{r_p}}_{\text{Bessel shift}}.
    \label{eq:diagonal}
\end{align}
The mode dependence is entirely confined to the diagonal through the centrifugal contribution $m^2/r_p$ in Eq.~(\ref{eq:diagonal}) --- the conservative and discrete counterpart of the Bessel-operator shift $m^2/r^2$ in $\mathcal{L}_m$ --- which grows with $m$ and diverges near the axis, progressively stiffening the diagonal for higher modes. The assembled system is
\begin{equation}
    \mathsf{L}_m\,\hat{\boldsymbol{\phi}}_m = \hat{\mathbf{s}}_m,
    \qquad
    \hat{\mathbf{s}}_m \equiv -\frac{1}{\varepsilon_0}\mathsf{R}\,\hat{\boldsymbol{\rho}}_m,
    \label{eq:linear_system}
\end{equation}
where $\mathsf{L}_m \in \mathbb{R}^{N_rN_z \times N_rN_z}$ assembles the stencil of Eq.~(\ref{eq:stencil}) and $\mathsf{R} = \mathrm{diag}(r_p)$ is the cylindrical Jacobian matrix, required for consistency on both sides of the discretised equation as a consequence of using the $r$-weighted conservative form within $\mathsf{L}_m$.

The on-axis boundary conditions specified in Eq.~(\ref{eq:axis_bc}) are incorporated directly into the finite-volume discretisation. At the cylindrical axis, $r_{1/2}=0$ defines the innermost radial cell face. For the axisymmetric mode ($m=0$), the regularity condition $\partial\hat{\phi}_0/\partial r=0$ is imposed as zero radial flux through the axis face. Consequently, the radial flux through the innermost face of the first radial control volume vanishes; the $\alpha_{p-1}$ radial coupling present in the interior stencil of Eq.~(\ref{eq:stencil}) is therefore absent from the axis-adjacent cell equation. For the non-axisymmetric modes ($m\geq 1$), the regularity condition $\hat{\phi}_m(0)=0$ is imposed as a Dirichlet condition by row substitution in the linear system: the corresponding matrix row is replaced by a corresponding identity row, yielding $1\times\hat{\phi}_m(0)=0$, with the corresponding entry of the right-hand-side vector $\hat{\mathbf{s}}_m$ set to zero.}

\subsection{Mode-Dependent Preconditioning and Multigrid Solver}
\label{sec:preconditioning}
 
{\color{black}The linear system in Eq.~(\ref{eq:linear_system}) is solved using a geometric
multigrid algorithm with Successive Over-Relaxation (SOR). As
demonstrated in Section~\ref{sec:modal_poisson}, the mode-dependent variation
in the conditioning of the discrete operator is confined to its diagonal
through the coefficient $A_{p,q}^{(m)}$. Consequently, a diagonal (Jacobi)
preconditioner provides an effective and inexpensive normalisation of the
system,
\begin{equation}
\mathsf{D}_A^{(m),-1}\mathsf{L}_m\hat{\boldsymbol{\phi}}_m
=
\mathsf{D}_A^{(m),-1}\hat{\mathbf{s}}_m,
\qquad
\mathsf{D}_A^{(m)}
=
\operatorname{diag}\!\left(A_{p,q}^{(m)}\right),
\label{eq:jacobi_precond}
\end{equation}
where the diagonal of the preconditioned operator is unity by construction.
The off-diagonal coefficients are correspondingly scaled by
$1/A_{p,q}^{(m)}$, reducing the variation in the operator spectrum across
Fourier modes. No factorisation is required; the coefficients $A_{p,q}^{(m)}$ are extracted from the assembled stencil once during initialisation and stored as a vector per mode.

Decomposing the preconditioned operator as
\begin{equation}
\mathsf{D}_A^{(m),-1}\mathsf{L}_m
=
\mathsf{I}
+
\tilde{\mathsf{L}}_m^{-}
+
\tilde{\mathsf{L}}_m^{+},
\end{equation}
where $\tilde{\mathsf{L}}_m^{-}$ and
$\tilde{\mathsf{L}}_m^{+}$ denote the strictly lower and upper triangular
parts, respectively, the SOR solver is written in affine form as
\begin{equation}
\begin{aligned}
\hat{\boldsymbol{\phi}}_m^{(k+1)}
&=
\left[
(1-\omega)\mathsf{I}
-\omega\tilde{\mathsf{L}}_m^{+}
\right]
\hat{\boldsymbol{\phi}}_m^{(k)}
-\omega\tilde{\mathsf{L}}_m^{-}
\hat{\boldsymbol{\phi}}_m^{(k+1)}
\\
&\quad
+\,
\omega\mathsf{D}_A^{(m),-1}\hat{\mathbf{s}}_m.
\end{aligned}
\label{eq:sor_affine}
\end{equation}

}{\color{black}The lower-triangular term is evaluated using the most recently updated
components of $\hat{\boldsymbol{\phi}}_m^{(k+1)}$, corresponding to a
forward-substitution sweep, while the upper-triangular term uses the
values from the previous iterate $\hat{\boldsymbol{\phi}}_m^{(k)}$.}{\color{black} $1<\omega<2$ is the relaxation parameter (setting $\omega=1$
recovers standard Gauss--Seidel). For modes $m\ge1$, the modal potential
$\hat{\boldsymbol{\phi}}_m$ is complex-valued whereas
$\mathsf{L}_m$ is purely real. The solver is therefore applied
independently to the real and imaginary components as two real-valued systems.

Coarse-grid operators are constructed by direct re-discretisation of
Eq.~(\ref{eq:modal_poisson}) on each coarsened mesh, preserving the
$m^2/r_p$ contribution throughout the multigrid hierarchy. The Jacobi
preconditioner is therefore recomputed trivially from the diagonal stencil
coefficients at every level.

The solver cost per mode is then $\mathcal{O}(2N_rN_z)$. Since the modal systems are mutually
independent, the total serial complexity of the field solve scales as
$\mathcal{O}((2N_m+1)N_rN_z)$. Ideal parallel execution permits all modal solves to proceed concurrently, reducing the wall-clock cost toward
$\mathcal{O}(2N_rN_z)$.

Verification of the second-order discretisation accuracy of the solver, using a Method of Manufactured Solutions (MMS) test, is provided in Appendix~\ref{app:mms}.}

\begin{figure*}[!t]
\centering
\includegraphics[width=0.5\textwidth]{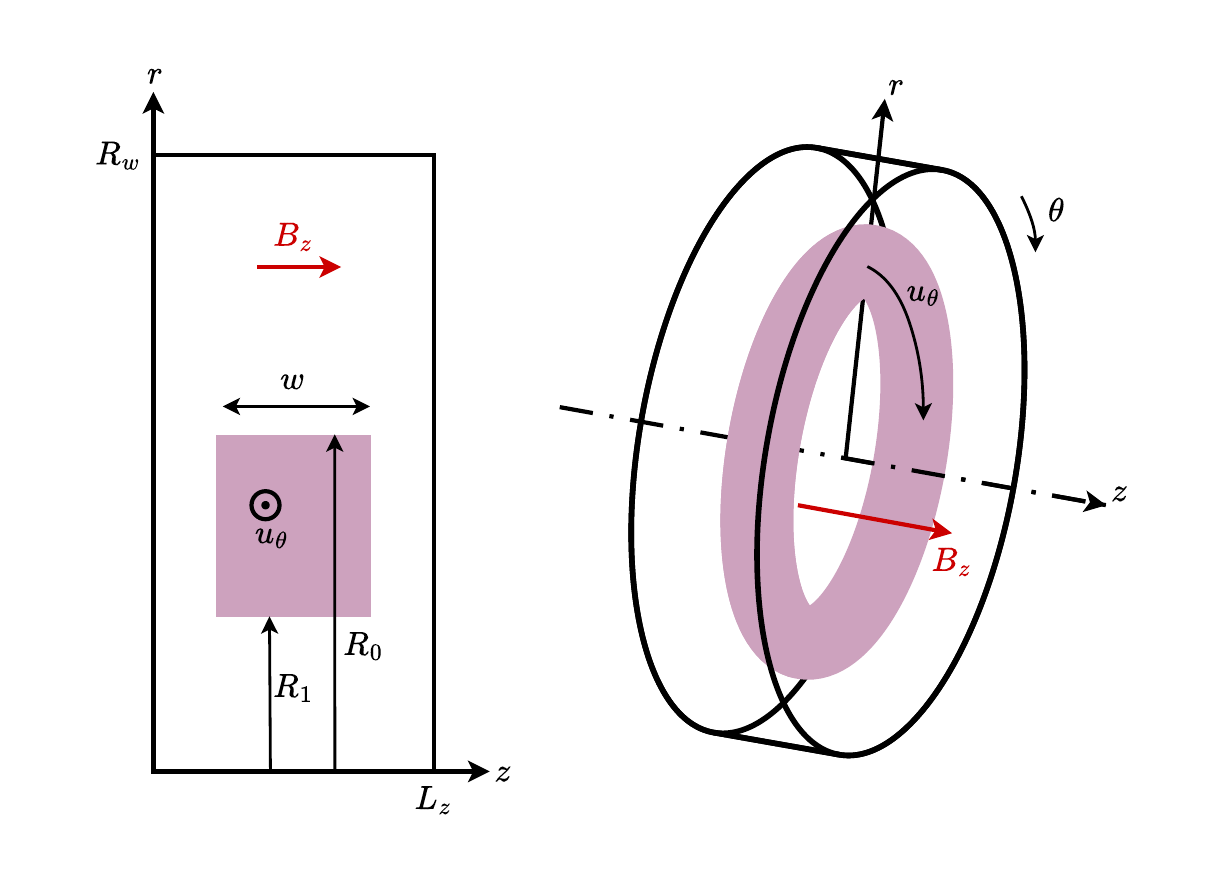}
\caption{Setup of the diocotron validation case.}
\label{fig:diocotron}
\end{figure*}

\section{Benchmarking}
\label{sec:benchmarking}

{\color{black}The validation strategy adopted here focuses on the underlying modal decomposition and reconstruction procedures rather than exhaustive demonstration of fully three-dimensional physics. First, the diocotron instability provides a case for which linear growth rates and eigenmode structures are known analytically, then the Landmark Penning discharge constitutes a widely used benchmark exhibiting strongly non-axisymmetric behaviour. Neither problem exercises substantial axial variation: the diocotron is axially uniform, and the Landmark Penning discharge is here resolved with a single axial cell $(N_z = 1)$ as a purely radial-azimuthal two-dimensional problem. Together they provide stringent tests of the modal field representation that forms the basis of the present method. The extension of the method to fully three-dimensional plasma cases, with significant axial dynamics, shall follow this work.}

\subsection{Diocotron Instability}


The diocotron instability is a purely electrostatic, shear-driven instability
of magnetised non-neutral or partially neutralised electron columns and
represents one of the most thoroughly characterised azimuthal instabilities in
plasma physics~\cite{davidson1990,Levy1964DiocotronCylindrical}.  It is the plasma analogue of the
Kelvin--Helmholtz instability in neutral fluids: free energy stored in the
radial gradient of the azimuthal $\mathbf{E}\times\mathbf{B}$ drift velocity
drives growing perturbations of the electron density column.  In the strongly
magnetised limit, where the electron Larmor radius is small compared with the
gradient scale length, electron fluid elements undergo purely azimuthal
$\mathbf{E}\times\mathbf{B}$ drift and the instability manifests as a growing
azimuthal ripple on the column boundary.

The instability is ideally suited as a primary validation benchmark for the
DECOMPIC method for several complementary reasons.  First, it is intrinsically
azimuthal: the linear eigenmode has no axial variation, so the quasi-3D field
decomposition captures the complete linear physics without any truncation error
from mode limitation.  Second, closed-form analytic predictions for both the
growth rate and oscillation frequency exist for hollow cylindrical equilibria,
providing unambiguous quantitative targets against which the simulation can be
judged.  Third, the non-linear phase — vortex roll-up, discrete vortex formation,
and eventual merger — is well documented and provides qualitative benchmarks
beyond the linear regime.  Finally, the test involves only electrons advancing
in a prescribed static magnetic field with a self-consistent electrostatic
field, making it a clean isolated test of the Fourier charge deposition, modal
Poisson solver, field reconstruction, and cylindrical Boris pusher, free from
the additional complexity of ion dynamics.

The equilibrium is a uniform hollow electron annulus of inner radius $R_1$,
outer radius $R_0$, axial width $w$, and uniform number density $n_e$ within the annulus,
surrounded by vacuum out to a conducting wall at $R_w$.  A stationary
neutralising ion background of density $n_{i} = n_e$ is confined to the same
annular region to ensure global quasi-neutrality.  A uniform axial magnetic
field $\mathbf{B} = B_0\hat{z}$ is prescribed.  The geometry is illustrated schematically in Figure~\ref{fig:diocotron}.

The characteristic frequency scale is the diocotron frequency
\begin{equation}
    \omega_D = \frac{\omega_{pe}^2}{2\Omega_{ce}}
             = \frac{n_e e}{2\varepsilon_0 B_0},
    \label{eq:omega_D}
\end{equation}
which sets the rate of rigid-rotor $\mathbf{E}\times\mathbf{B}$ precession of
the electron column and is used to normalise time throughout this section,
$t_D \equiv \omega_D t$.

Linear theory for azimuthal mode $m$ gives a complex eigenfrequency
$\omega_m = \omega_{r,m} + i\gamma_m$ with an imaginary part that yields the growth rate~\cite{davidson1990},

{\color{black}
\begin{align}
\frac{\gamma_m}{\omega_D}
&= \left\{
\begin{aligned}
&\left(\frac{R_1}{R_0}\right)^{2m}
\left(
1-\left(\frac{R_0}{R_w}\right)^{2m}
\right)^2
-\frac{1}{4}
\Bigg[
2-\\
&\left(\frac{R_1}{R_w}\right)^{2m}
-\left(\frac{R_0}{R_w}\right)^{2m}
-m\left(
1-\left(\frac{R_1}{R_0}\right)^2
\right)
\Bigg]^2
\end{aligned}
\right\}^\frac{1}{2}
\label{eq:dioc_growth_norm}
\end{align}
}

\begin{figure*}[!t]
\centering
\includegraphics[width=\textwidth]{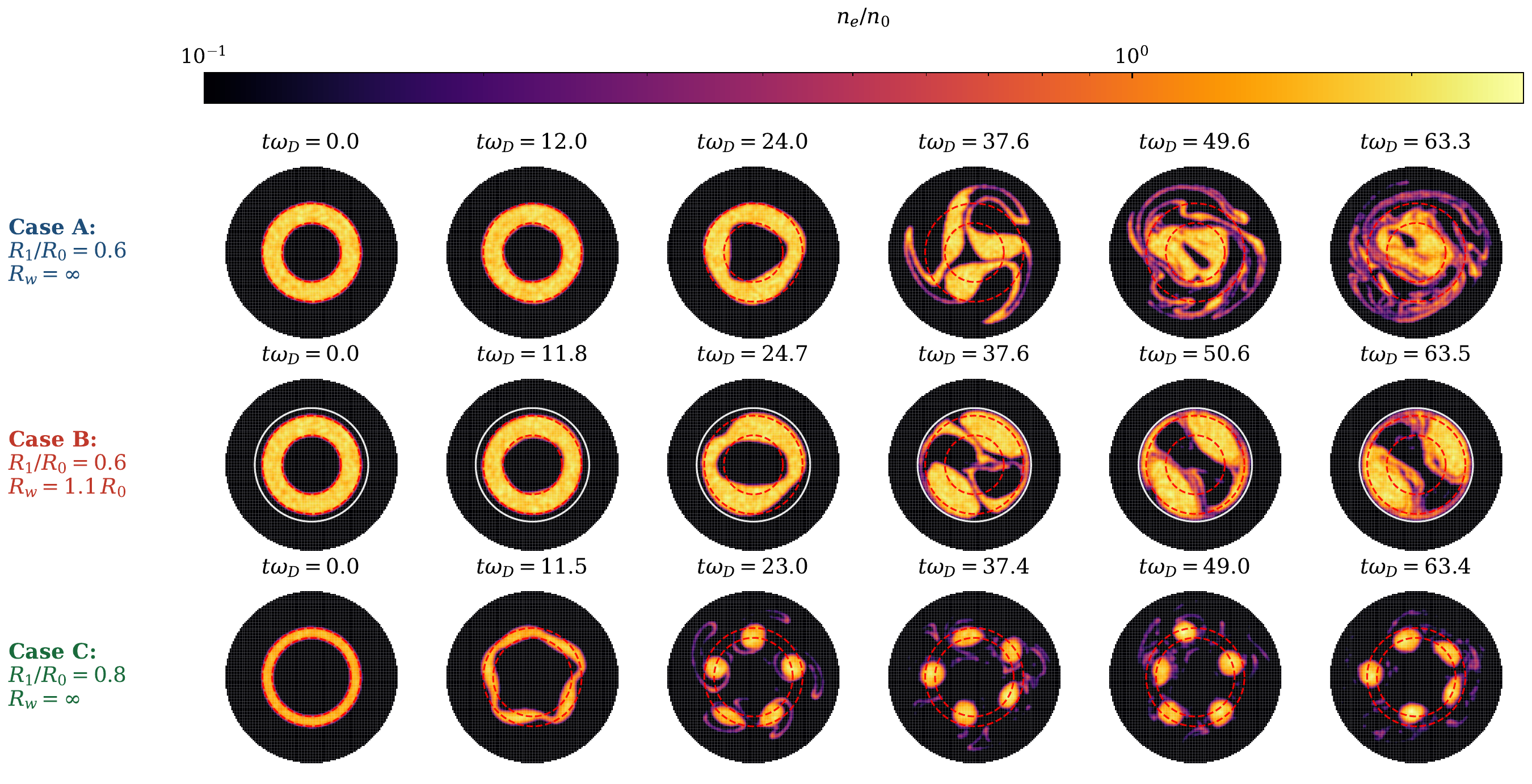}
\caption{Time evolution of the normalised electron density distribution for Cases A, B and C. The initial annulus is shown in red. The outer wall is shown in white in Case B.}
\label{fig:diocotron_validation}
\end{figure*}

Here the diocotron frequency $\omega_D$ defined in Eq.~(\ref{eq:omega_D}) has
absorbed the density and field dependence, so Eq.~(\ref{eq:dioc_growth_norm}) depends only on the three geometric ratios $R_1/R_0$, $R_0/R_w$, and the mode number $m$.  In the limit $R_w \to \infty$
the wall term $(R_0/R_w)^{2m}\to 0$ and both expressions simplify; bringing the
wall closer stabilises the instability by reducing the effective shear available
to the mode.  The growth rate is maximised at an intermediate wall proximity,
and the most unstable mode shifts to higher $m$ as the annulus becomes thinner
($R_1/R_0 \to 1$). Eq. (\ref{eq:dioc_growth_norm}) can also be used to identify the most unstable dominant mode m$^*$ --- that with the highest growth rate --- for any given geometry.

Three geometrically distinct configurations are studied, spanning different
combinations of annular thickness and wall proximity, to assess the code across
a range of mode numbers and growth rates.  Their normalised parameters are
collected in Table~\ref{tab:diocotron_cases}.  In all cases the physical density
and magnetic field enter only through $\omega_D$ and the time normalisation;
the dynamics are governed entirely by the geometric ratios.

The initially injected electron population is Maxwellian at 1 eV with density $n_e=1\times 10^{12}$ m$^{-3}$. The applied magnetic field is $B_z= 25$ G. The stretched radial mesh consists of 256 cells and strictly adheres to the constraints listed in Section III. {\color{black}The outer radius, which serves as the radial geometrical reference, is $R_0 = 50\lambda_D$}. The axial domain length $L_z$ is limited to $\lambda_D$ and resolved by 10 cells, with the annulus width $w=0.2\lambda_D$; this limits the effects of axial dispersion that would distort the desired pure assessment of the azimuthal instability. The limiting time-step is $0.1\omega_{pe}^{-1}$, with the simulations run to {\color{black}$t\omega_{pe}=1000$} which corresponds to approximately $t\omega_D=64$. Nominally, the average number of particles per cell $N_{ppc}=200$ (which leads to $N_{ppc}>600$ in the regions occupied by plasma) and the mode number is truncated at $N_m = 3m^{*}+1$ to ensure higher harmonics of the dominant mode are retained. {\color{black}The wall at $R_w$ is grounded at $\phi_w=0$ and all particles there are absorbed. In the case A and C instances of a wall at infinity, the geometric ratio is numerically treated as $R_w/R_0=10$.}

\begin{table}[h]
\centering
\caption{Normalised input parameters for the three diocotron test cases. The most unstable mode $m^*$ and the analytically predicted normalised growth rate
$\gamma_{m^*}/\omega_D$ from Eq.~(\ref{eq:dioc_growth_norm}) are given for each
case. $N_m$ is the number of azimuthal Fourier modes retained.}
\label{tab:diocotron_cases}
\begin{tabular}{lccccc}
\hline
\textbf{Case} & $R_1/R_0$ & $R_w/R_0$ & $m^*$ & $\gamma_{m^*}/\omega_D$ & $N_m$ \\
\hline
A & 0.6 & $\infty$ & 3 & 0.212 &  10  \\
B & 0.6 & 1.1 & 2 & 0.111  & 7  \\
C & 0.8 & $\infty$ & 5 & 0.312  & 16 \\
\hline
\end{tabular}
\end{table}



The reconstructed radial-azimuthal electron densities for all three cases are presented together in Figure~\ref{fig:diocotron_validation}. This reconstruction is achieved via Eq. \ref{eq:rho_series} using a resolution of 256 azimuthal points. The presented snapshots trace the full lifecycle of the instability from destabilisation, vortex roll-up, discrete vortex formation, and eventual merger. The unperturbed annulus develops an $m^*$-periodic corrugation at the outer boundary during the linear phase, with three, two, and five lobes for Cases A, B, and C respectively.  Non-linear steepening causes the lobes to roll up into discrete electron vortices that orbit the column axis, qualitatively reproducing well-documented diocotron behaviour. The last two snapshots of Case A then illustrate the expected merger of the vortices. In Case B, bringing the conducting wall to $R_w = 1.1R_0$ suppresses radial excursions and the fragmentation is less complete, consistent also with the reduced growth rate. In Case C, the thinner annulus decreases the gradient length scale leading to faster onset and growth of the instability. In the final snapshot of Case C, the commencement of vortex merger can be seen in the asymmetry of the 5 orbiting discrete vortices.

\begin{figure*}[!t]
\centering
\includegraphics[width=\textwidth]{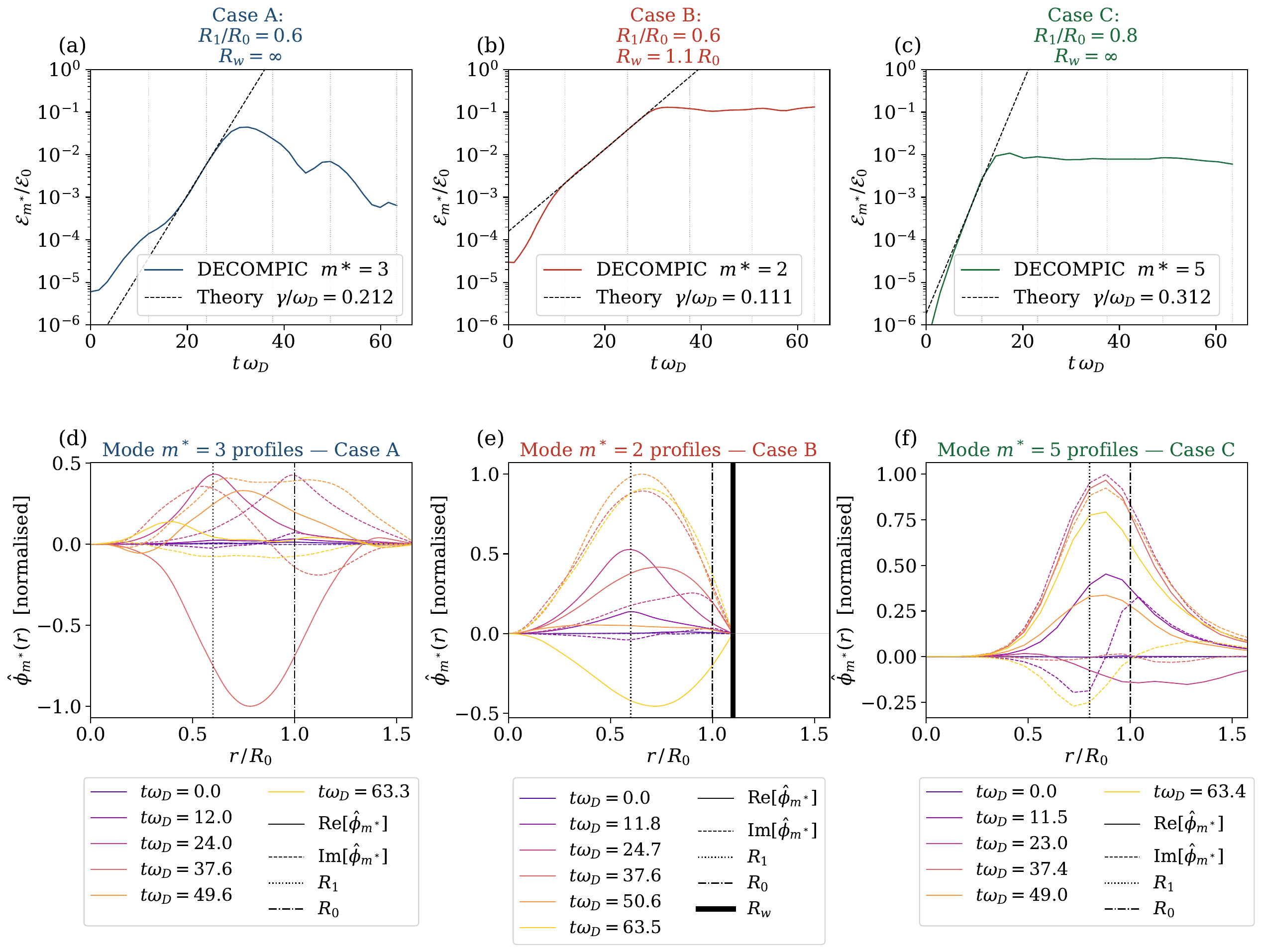}
\caption{(a)-(c): Time evolution of the normalised dominant-mode field energy for Cases A, B and C, shown with the theoretical growth; grey vertical lines correspond to the time snapshots given in Figure \ref{fig:diocotron_validation}. (d)-(f): Radial profiles of the dominant-mode Fourier coefficients at the corresponding times.}
\label{fig:diocotron_validation_profiles}
\end{figure*}

The evolution of the dominant-mode field energy $\mathcal{E}_{m^*}(t) = (\varepsilon_0/2)\int|\hat{\mathbf{E}}_{m^*}|^2\,dV$ is given in Figures \ref{fig:diocotron_validation_profiles} (a)-(c).  In each case a sustained interval of clean exponential growth is observed, during which $\mathcal{E}_{m^*}\propto e^{2\gamma_{m^*}t}$, consistent with linear amplification of the electric field amplitude. The theoretical growth curves from Eq.~(\ref{eq:dioc_growth_norm}) are then overlaid. To quantify the agreement, the simulation growth rate is extracted from each simulation by fitting $\ln\mathcal{E}_{m^*}(t)$ over the linear phase via least-squares regression, yielding a slope of $2\gamma$. The relative errors in growth rate were found to be 3.4\%, 1.1\%, and 7.1\% for cases (a), (b), and (c) respectively, confirming excellent quantitative agreement throughout the linear phase. Consistent with the vortex merging behaviour observed in Figure~\ref{fig:diocotron_validation}, the $m^*=3$ mode in Case A can be seen to experience a post-saturation sustained damped oscillation, during which the energy rapidly decays.

Panels (d)-(f) show the radial distribution of the normalised eigenpotential coefficients $\hat{\phi}_{m^*}(r)$ at times corresponding to the snapshots of Figure \ref{fig:diocotron_validation}.  Throughout the initial time, the real and imaginary parts of the dominant-mode retain a nearly time-invariant profile --- growing in amplitude but not in structure --- with the imaginary part spatially shifted by circa quarter-wavelength in the azimuthal direction relative to the real part, as required for a mode rotating at the precession frequency $\omega_{r,m^*}$. This phase relationship confirms that the mode is a unidirectional rotating wave rather than a standing oscillation, consistent with the complex exponential structure expected of the diocotron eigenfunction.  The potential peaks at the annulus boundaries where the $\mathbf{E}\times\mathbf{B}$ shear is largest, confirming the correct localisation of the eigenpotential behaviour.  Broadening of the profiles and the emergence of fine radial structure mark the onset of non-linearity, most evident in Case C owing to its higher growth rate. While the eigenpotential structure encodes considerable diagnostic insight into the diocotron instability, a systematic analysis is beyond the scope of this validation study and is reserved for future investigation.


\begin{figure*}[!t]
\centering
\includegraphics[width=0.8\textwidth]{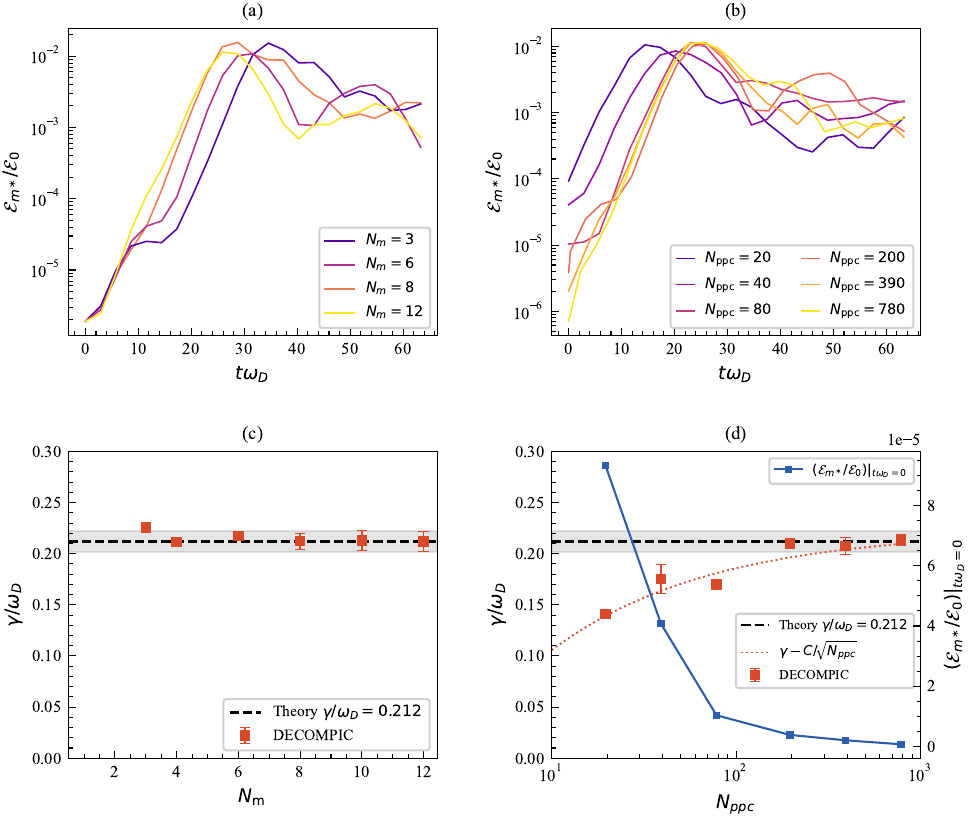}
\caption{Time-evolution of the normalised $m^*=3$ mode energy for Case A with variation in: (a) the truncated maximum mode number $N_{m}$; (b) the macroparticle count per cell $N_{ppc}$. (c) and (d) provide corresponding recovered growth rate with increasing $N_{m}$ and $N_{ppc}$ respectively against the theoretical (displayed with a $\pm$5\% errorbar). The evolution of the initial mode energy noise floor is also shown in (d).}
\label{fig:diocotron_convergence}
\end{figure*}

Having established that DECOMPIC reproduces the correct diocotron physics, next is to examine the dependence of the extracted growth rate on the two principal numerical parameters: The truncated mode number $N_m$ and the macroparticle count $N_{ppc}$.  All tests use Case A ($R_1/R_0=0.6$, $R_w\to\infty$, $m^*=3$, $\gamma_\mathrm{th}/\omega_D=0.212$) and are summarised in Figure~\ref{fig:diocotron_convergence}. The key result of the $N_m$ sweep in panel (c) is that the theoretical linear growth rate is recovered within 5\% at all $N_m$. The rate peaks at $N_m = 3$ with $\gamma/\omega_D$=0.223, where the dominant mode is resolved but it is possible no higher harmonics are available to drain energy from it, resulting in a slightly elevated apparent growth rate. The shifted onset times and differing peak energies visible in the energy histories of panel (a) are expected, since fewer retained modes mean fewer non-linear coupling channels available to receive energy from $m^*=3$, naturally altering the saturation and pre-linear phases without affecting the underlying linear physics of the growth phase. The $N_{ppc}$ convergence of panel (d) shows a different character: the extracted growth rate exhibits a systematic downward bias at low {\color{black}$N_{ppc}$}, following the typically expected {\color{black}$\sqrt{N_{ppc}}^{-1}$} scaling as shown, which arises because shot noise provides a large spurious noise seed above or comparable to the physical signal of the growing mode, illustrated also in panel (d) via the descending initial spectral energy, that acts to drive the mode toward non-linear saturation before a clean linear phase is established, may artificially cause mode-coupling that removes energy from $m^*=3$, whilst also fundamentally inhibiting the signal due to reduced SNR. As $N_{ppc}$ increases the noise floor recedes, the linear phase lengthens, and the extracted rate converges to the theoretical value from below; $<$5\% error is achieved at $N_{ppc}>200$ and convergence is seen within the linear phase of the mode energy histories in panel (b). Together, the near-independence in $N_m$ and convergence in $N_{ppc}$ demonstrate that the DECOMPIC implementation is free from systematic numerical artefacts: the dominant error source is statistical shot noise, which is well understood within any PIC method, scales predictably with $N_{ppc}$, and vanishes in the limit of large particle count.

\subsection{Penning Discharge Benchmark}

\begin{figure*}[!t]
\centering
\includegraphics[width=\textwidth]{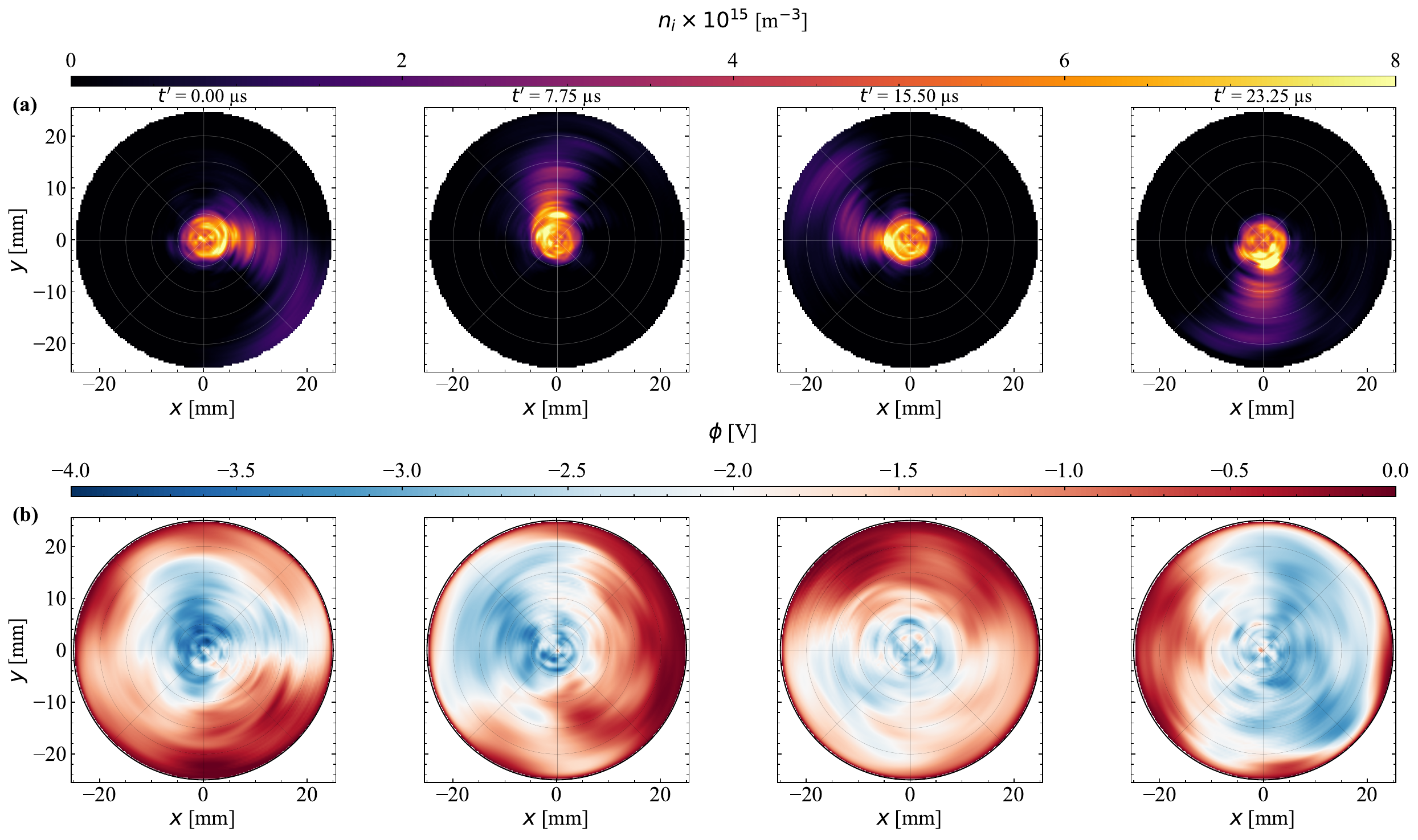}
\caption{Contour plots of (a) ion density $n_i$ and (b) plasma potential $\phi$ at four different phases of spoke rotation during the quasi-steady state. Times $t'$ are given relative to the spoke in the 90$^\circ$ position.}
\label{fig:spoke_snapshots}
\end{figure*}

Recently, a new benchmark for two-dimensional partially-magnetised $E \times B$ plasmas has been introduced by the \textit{Landmark} project \cite{landmark_penning}. This benchmark considers a collisionless helium-4 Penning discharge and was simulated by seventeen PIC codes in a collaborative community effort. The emergence of large-scale coherent structures in the form of rotating plasma spokes endows this configuration with an enormous range of time scales, making it particularly challenging to reproduce. It is generally understood that a collisionless Simon-Hoh instability (CSHI) occurs due to steep gradients at the plasma edge and saturates into the rotating spoke \cite{Charoy2020AnomalousTransport}.  The configuration is a close derivative of the previous Diocotron case; the Penning discharge considers a quasi-neutral ion-electron plasma column instead of a non-neutral electron annulus. 

\begin{figure*}[!t]
\centering
\includegraphics[width=\textwidth]{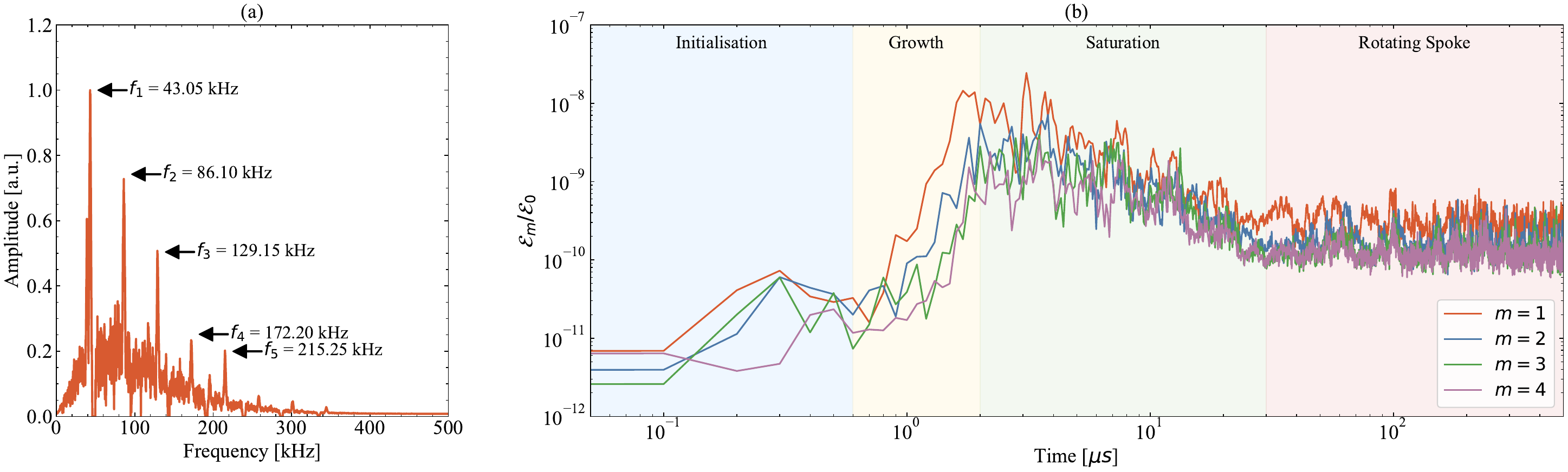}
\includegraphics[width=0.75\textwidth]{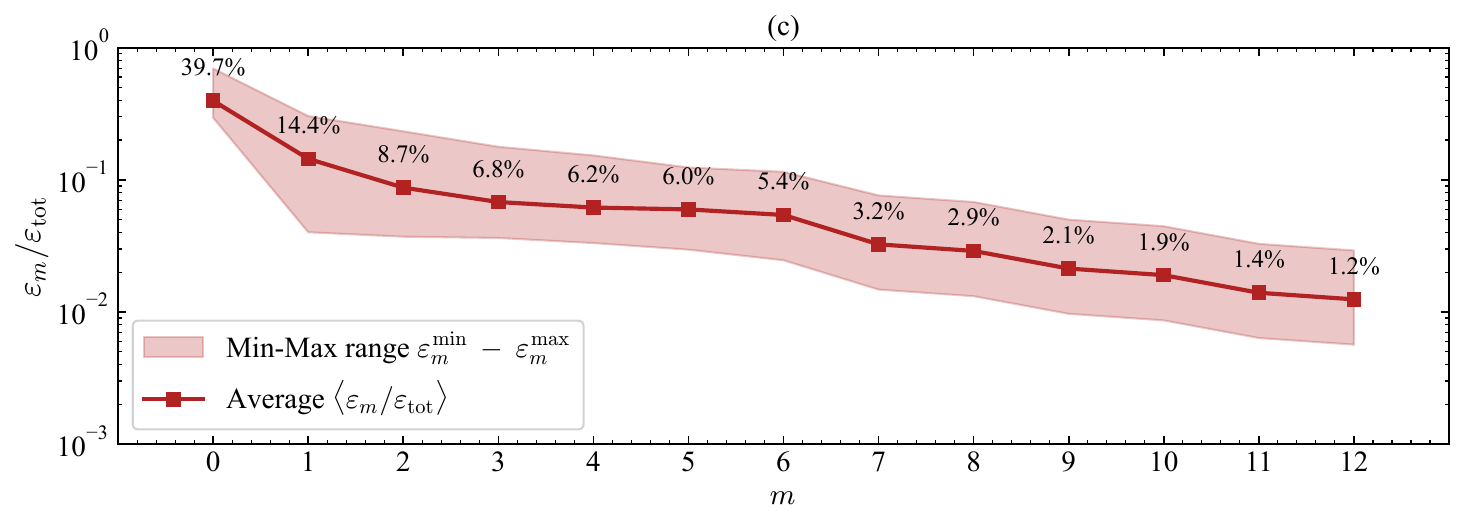}
\caption{(a) Frequency spectrum of the virtual probe ion density signal during the 200-500 $\mu s$ steady-state; (b) Temporal evolution of the first four mode energies (m = 1–4), illustrating the progression from initialisation through linear growth and non-linear saturation into a steady rotating spoke regime; (c) Modal global energy spectrum during the 200-500 $\mu s$ steady-state.}
\label{fig:fft}
\end{figure*}

\begin{figure*}[!t]
\centering
\includegraphics[width=\textwidth]{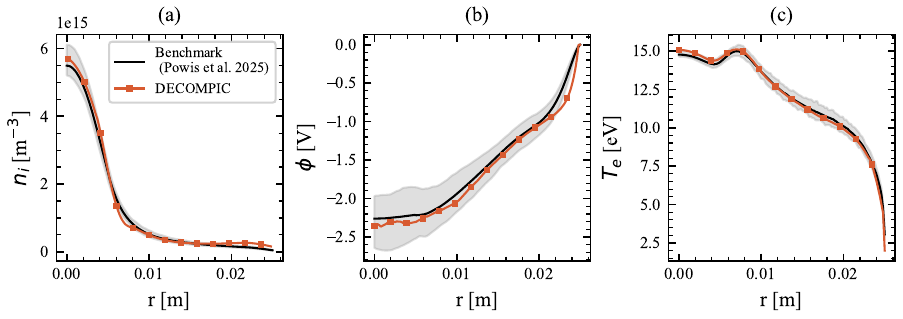}
\caption{Time-averaged radial profiles of (a) ion density $n_i$, (b) plasma potential $\phi$, and (c) electron temperature $T_e$ from DECOMPIC, compared against the Landmark benchmark~\cite{landmark_penning}. The uncertainty band spans the minimum and maximum of a rolling 300 $\mu$s average over 7000 windows. DECOMPIC profiles are averaged over the interval 200 - 500 $\mu$s of the simulation.}
\label{fig:radialbenchmark}
\end{figure*}

The simulation domain of the Penning discharge case is therefore the same as that of the Diocotron case illustrated in Figure \ref{fig:diocotron}, except $R_i = 0$. The wall radius $R_w = 25$ mm and is grounded with potential $\phi_w = 0$ V; all particles reaching the wall are absorbed. It is critical to note that the \textit{Landmark} case relaxed the curved geometry of the Penning discharge experiment to a Cartesian 50 mm square, but that the implementation in DECOMPIC is cylindrical-axisymmetric. {\color{black}The total simulation time is 500 $\mu s$ at a time-step of $40$ ps, in strict adherence to the benchmark; for the quasi-steady state peak density of circa $6\times 10^{15}~m^{-3}$ (see Fig. \ref{fig:radialbenchmark} (a)), this corresponds to $\omega_{pe}\Delta t\approx 0.18$, which lies within the commonly used range of stability coefficient for explicit electrostatic PIC simulations.} A uniform, constant magnetic field $B = 100$ G is externally applied in the axial direction. A virtual measurement probe is placed within the simulation domain at the half-radius $r = 12.5$ mm, $\theta = 0$. The domain is initially empty. A cylindrical particle injection source with the radius $R_o = 5$ mm then injects helium ions and electrons at fixed current density throughout the simulation time. The injected electrons and ions are sampled from Maxwellian distribution functions at temperatures of 15 eV and 0.025 eV respectively. Full details on the physical and numerical conditions of the benchmark problem can be found in the \textit{Landmark} reference \cite{landmark_penning}. 

{\color{black}The benchmark employs a 256$\times$256 uniform mesh, yielding $N_xN_y = 65536$ coupled unknowns in the Poisson solve at every time-step. DECOMPIC replaces this two-dimensional Cartesian solve with $N_m+1$ independent one-dimensional radial solves of length $N_r$, where $N_r$ corresponds to only the half-domain from the axis to the wall due to the transition from Cartesian to axisymmetric coordinates (for the purposes of this benchmark, the axial extent is limited to a single cell $N_z=1$). {\color{black}The demonstrated reduction here is therefore from a single two-dimensional $(r,\theta)$ problem to $N_m+1$ independent one-dimensional problems}. For the present configuration with $N_r=128$ and $N_m=12$, this gives $(N_m+1)N_r = 1664$ total independent modal potential unknowns against 65536 --- a reduction of approximately 39$\times$. Since modes with $m\geq1$ are complex and are represented by separate real and imaginary components in the present solver implementation, the equivalent count in terms of real-valued solver degrees of freedom is $(2N_m+1)N_r = 3200$, corresponding to a reduction of approximately 20$\times$. The former count reflects the reduction achieved by the DECOMPIC formulation, while the latter represents the scalar degree-of-freedom count for the specific real-valued numerical implementation used here.} 

Since the number of macroparticles required to achieve a given statistical resolution also scales with the number of cells through the particles-per-cell constraint, the particle population is reduced by a similar factor. At the benchmark macroparticle weight of $W=10^5$ this corresponds to a reduction from approximately $10^6$ to $\mathcal{O}(10^5)$ macroparticles. These reductions represent a commensurate saving in total wall-clock time, therefore bringing the full Penning discharge simulation from a multi-hundred-core high-performance computing allocation to a run completable on a single multi-core workstation.

{\color{black}The choice of $N_m = 12$ was made a priori to retain the dominant $m = 1$ spoke together with a sufficient range of higher-order azimuthal harmonics and smaller-scale structures. Rotating spokes in E$\times$B discharges are generally dominated by the $m = 1$ mode, while higher-order $m > 1$ structures may coexist with the spoke and contribute to smaller-scale azimuthal dynamics. For the characteristic discharge radius of 5 mm considered here, $m = 12$ corresponds to an azimuthal wavelength of approximately 2.6 mm, approximately 10\% of the 25 mm computational-domain radius. This was considered sufficient to resolve the characteristic azimuthal structure of the discharge while retaining the computational benefits of the modal decomposition. The resulting modal-energy spectrum of Figure 8 (c) provides an a posteriori assessment of the adequacy of this truncation, as well as the absence of Gibbs aliasing in Figure~\ref{fig:spoke_snapshots}}.

Figure~\ref{fig:spoke_snapshots} (a) shows four snapshots of the reconstructed ion number density $n_i$ separated by approximately one quarter of the spoke rotation period, sampled after the discharge has reached quasi-steady state ($t > 200$~$\mu$s), referenced from the left-most snapshot as $t'=0$, using 256 azimuthal reconstruction points and mapped to equivalent Cartesian coordinates for clarity. {\color{black}Since the modal coefficients are obtained from particle deposition, they satisfy the Hermitian symmetry condition $n_{i,-m}=n_{i,m}^{*}$, ensuring a real-valued reconstructed density. It is noted, however, that positivity is not mathematically guaranteed after truncation of a Fourier representation, since finite-mode reconstructions can introduce small oscillatory deviations analogous to Gibbs spectral ringing. The reconstructed ion density was monitored during the simulations and no non-physical negative densities were observed. The appearance of widespread negative reconstructed densities would indicate that the retained azimuthal mode set up to $N_m$ is insufficient to accurately represent the angular structure.}

The spoke manifests as a single elongated high-density arm rotating counter-clockwise---in the direction of the electron $\mathbf{E}\times\mathbf{B}$ drift---while the plasma outside the spoke arm and the central injection region is nearly evacuated.  This structure is qualitatively identical to that shown in Figure~4 of the benchmark paper~\cite{landmark_penning}, where the same vacuum character of the inter-spoke region is highlighted as a defining feature of the discharge and a primary source of diagnostic difficulty.

The four quarter-period snapshots illustrate that the spoke rotation is not rigid: the arm narrows and broadens as it sweeps, and transient density filaments are shed from its trailing edge, consistent with the persistent fine-scale structures noted in the benchmark and attributed to ion-acoustic or electron-cyclotron-drift instabilities excited along the spoke arm~\cite{landmark_penning}. Panel (b) shows the corresponding plasma potential $\phi$ reconstructed from Eq. (\ref{eq:phi_proj}). A rotating potential well of around -2 to -4 V leads the density spoke by approximately $\pi/2$ in phase, providing the azimuthal electric field that ultimately pulls the ions in the rotation. 

{\color{black}It is prudent to note that the spoke in Figure~\ref{fig:spoke_snapshots}(a) exhibits some qualitative differences from the benchmark~\cite{landmark_penning}, most notably a broader and less sharply localised azimuthal density structure. This may partly reflect the strongly chaotic nature of the spoke dynamics, as identified by Powis et al.~\cite{landmark_penning}, such that instantaneous snapshots are not expected to be identical. The finite azimuthal modal representation may also introduce the aforementioned Gibbs-type aliasing at sharp gradients along the spoke-arm edge, affecting the instantaneous appearance without necessarily altering the time-averaged quantitative properties. Differences in snapshot timing and rendering likely contribute as well. Notably, a similarly broadened spoke appearance is also visible in the reduced-order PIC results of Reza et al.~\cite{Reza2024PenningAxialRadial} for the same benchmark, suggesting that this observation is not necessarily specific to the present method.}

Thirteen complete rotations were captured within a 300~$\mu$s averaging window from 200-500 $\mu s$, matching that observed in the 500~$\mu$s PPPL reference run of reference~\cite{landmark_penning}.  The mean spoke rotation frequency extracted by peak-finding via the virtual probe was found to be $f_\mathrm{spoke} \approx 43.05$~kHz, within the benchmark-determined uncertainty bounds of $41.1$--$46.1$~kHz centred on the long-time reference value of 43.2~kHz (measured over 163 rotations by the PPPL code over 4~ms)~\cite{landmark_penning}. Cycle-to-cycle variation across the 13 observed rotations spanned a range of approximately $42.45$--$44.82$~kHz, comparable to the benchmark spread (41.1--46.1~kHz over sequential groups of 13 rotations), confirming that the chaotic variability of the spoke period is reproduced. The temporal Fourier transform of the probe signal is given in Figure \ref{fig:fft} (a) and shows the dominant peak at $f_\mathrm{spoke}$ and clearly resolved
harmonics at $2f_\mathrm{spoke}$ to $5f_\mathrm{spoke}$, with spectral power decaying by roughly one order of magnitude per harmonic.  The rapid decay of harmonics is consistent with a nearly sinusoidal character of the probe trace modulated by the rotating coherent structure; the finite harmonic content reflects the non-rigid, non-uniform shape of the spoke arm.

\begin{figure*}[!t]
\centering
\includegraphics[width=0.8\textwidth]{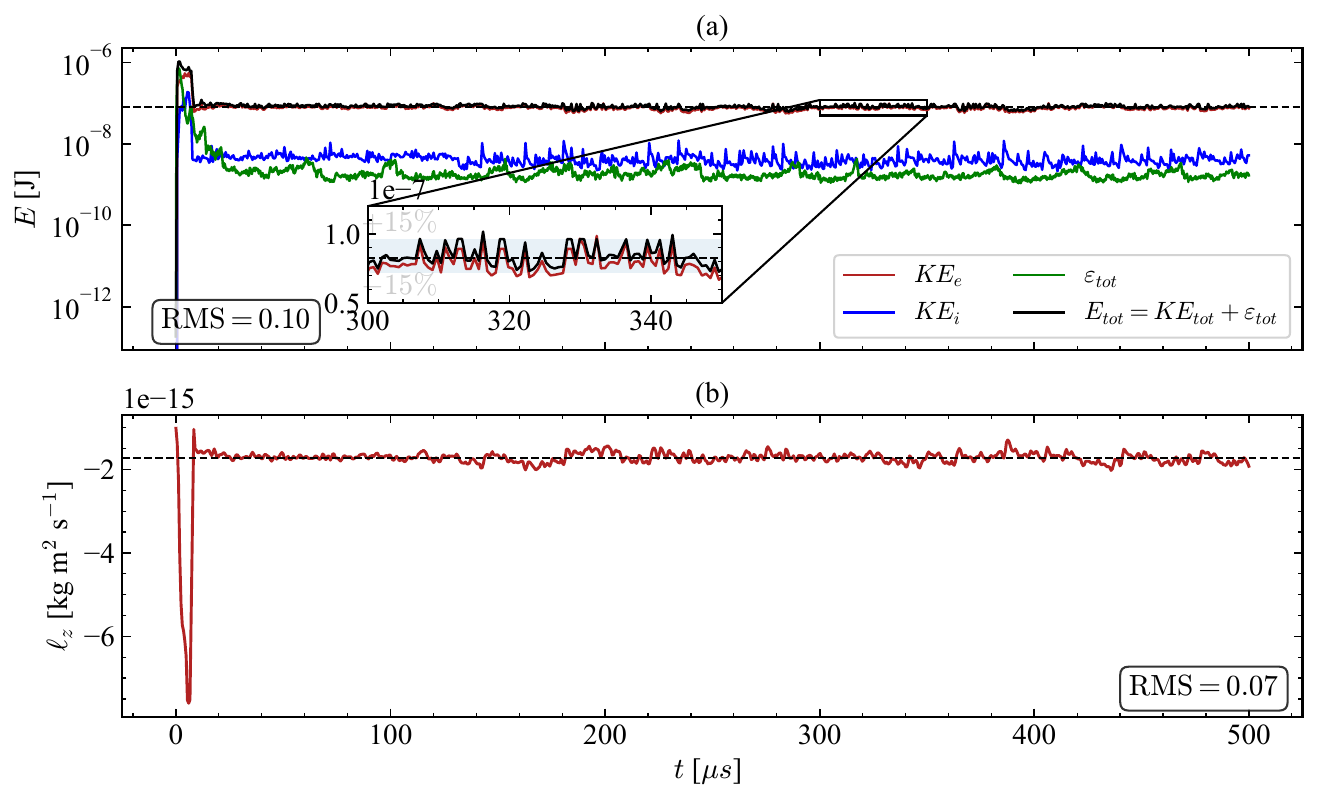}
\caption{Time histories of (a) the total system energy, comprising particle kinetic and total electrostatic field energy; and (b) the total particle axial angular momentum.}
\label{fig:conservation}
\end{figure*}

The dominant dipolar structure of the spoke, via the $m=1$ mode, is captured by Figure~\ref{fig:fft} (b), which shows the time evolution of the electrostatic field energy carried by modes $m=1-4$ as a proportion of the mean-field energy of $m=0$. Modes $m > 4$ are omitted for clarity, since they exist at amplitudes similar to the noise floor. The $m=1$ energy exhibits a prolonged exponential growth phase beginning around $t \approx 20$~$\mu$s, once sufficient plasma density has accumulated to render the CSHI unstable.  Non-linear saturation is reached at $t \approx 150$--$200$~$\mu$s, after which $\mathcal{E}_1$ oscillates quasi-periodically at the spoke rotation frequency, reflecting the cyclic intensification and partial dispersal of the arm. The $m=2$ mode follows with a slight delay and a shallower growth slope, playing the dual role of an
independently unstable mode (with a CSHI growth rate approximately twice that of $m=1$) and of the first non-linear harmonic generated by quadratic self-interaction of $m=1$.  At saturation, $\mathcal{E}_2$ lies approximately one to two orders of magnitude below $\mathcal{E}_1$ and oscillates at twice the spoke frequency, confirming the harmonic character.  Modes $m=3-4$ follow the same pattern at progressively later onset times and lower saturated amplitudes. A slow modulation of all mode energies at timescales longer than the spoke period is visible in the quasi-steady-state portion of the histories.  This low-frequency envelope is consistent with the intrinsically chaotic, multiscale character of the discharge.

{\color{black}Figure~\ref{fig:fft} (c) presents the modal global energy spectrum during the steady-state 200–500 $\mu s$ phase, which exhibits an expected overall decay in energy with increasing azimuthal mode number and a min-max range indicating the anticipated temporal variability. There is a modest local plateau at $m = 4-6$; this may arise  from non-linear harmonic content or mode coupling associated with the spoke structure. It may also indicate smaller-scale sub-dominant instability modes developing within or around the spoke arm, such as an ion-acoustic or ECDI-type instability, especially seeing as no deviation from typical modal decay is observed in the spoke rotation spectrum of Figure~\ref{fig:fft} (a). Since the modal energy spectrum represents a global measure of the energy associated with each azimuthal Fourier mode, it does not provide direct information on the spatial localisation or local physical origin of such possible individual modal structures. A detailed investigation of the physical mechanisms underlying the higher-order modal content, including possible secondary instabilities within the spoke, is beyond the scope of the present validation study and is therefore left for future work. The overall decay of the modal energy towards the truncation limit indicates that the retained modes capture the prevailing energetic content, with >90\% of the field energy contained within modes $m = 0-7$. The modes approaching the truncation limit therefore contribute comparatively little to the total field energy, with the $m = 12$ mode carrying only 1.2\% of the average total field energy, supporting the adequacy of the $N_m = 12$ truncation for the present simulation.}

\begin{table*}[t]
\centering
\caption{Approximate wall-clock runtimes and CPU-hour equivalents for the 500~$\mu$s Landmark
Penning discharge benchmark. Benchmark data from Table~2 of Powis et al.\cite{landmark_penning}.}
\label{tab:cpu_comparison}
\footnotesize
\setlength{\tabcolsep}{4pt}
\renewcommand{\arraystretch}{1.1}
\resizebox{\textwidth}{!}{
\begin{tabular}{lcccccc}

\hline
\hline

Code/Institute &
Language &
Poisson Solver &
Parallelism &
Hardware &
Wall-clock time (h) &
CPU-hours (h)  \\
\hline
\hline

LTP-PIC (PPPL) &
C/C++ &
Hypre &
Particle MPI + OpenMP + OpenACC &
4$\times$ GPU + 4$\times$ CPU cores &
82 &
-- \\
\hline

EP-PIC2D (Laplace) &
Fortran &
PARDISO &
Particle MPI + OpenMP &
72$\times$ CPU cores &
71 &
5112 \\
\hline

LePIC2D (Laplace) &
Fortran &
in-house multigrid &
Particle MPI + OpenMP &
28$\times$ CPU cores &
128 &
3584 \\
\hline

RUBPIC &
C/C++ &
NAG geometric multigrid &
Hybrid CUDA &
1$\times$ GPU + 1$\times$ CPU core &
80 &
-- \\
\hline

MZ/X-PIC (Wigner) &
C/C++ &
in-house spectral &
Particle CUDA &
1$\times$ GPU + 1$\times$ CPU core &
240 &
-- \\
\hline

DCU &
C &
in-house multigrid &
Domain OpenMP &
32$\times$ CPU cores &
240 &
7680 \\
\hline

RHEI (ONERA) &
Fortran &
in-house multigrid &
Hybrid OpenMP + MPI &
48$\times$ CPU cores &
284 &
18432 \\
\hline

PICASO (UC3M) &
Fortran &
PARDISO &
Particle OpenMP &
40$\times$ CPU cores &
181 &
7240 \\
\hline

PICASO-G (UC3M) &
Python &
in-house spectral &
Particle CUDA &
1$\times$ GPU &
10 &
-- \\
\hline

EDIPIC-2D (USask) &
Fortran &
PETSc + Hypre &
Hybrid MPI &
256$\times$ CPU cores &
1008 &
258048 \\
\hline

LPPic &
Fortran &
Hypre &
Domain MPI &
128$\times$ CPU cores &
1320 &
168960 \\
\hline

XPIC2D (Stanford) &
C/C++ &
Hypre &
Particle MPI &
32$\times$ CPU cores &
888 &
28416 \\
\hline

PANTERA (VKI) &
Fortran &
PETSc &
Hybrid MPI &
768$\times$ CPU cores &
72 &
55296 \\
\hline

AVIP-PIC (CERFACS) &
Fortran &
PETSc &
Domain MPI &
192$\times$ CPU cores &
185 &
35520 \\
\hline

PICCOLO (CNR-ENEA) &
Fortran &
PETSc + Hypre &
Hybrid MPI + OpenMP &
42$\times$ CPU cores &
720 &
30240 \\
\hline

Aleph (Sandia) &
C/C++ &
Trilinos &
Hybrid MPI &
640$\times$ CPU cores &
336 &
215040 \\
\hline
\hline

\textbf{DECOMPIC} &
\textbf{Java} &
\textbf{in-house multigrid} &
\textbf{Hybrid Java Fork/Join} &
\textbf{16$\times$ CPU cores} &
\textbf{40} &
\textbf{640} \\
\hline
\hline

\end{tabular}
}
\end{table*}

Figure~\ref{fig:radialbenchmark} compares the time-averaged azimuthally-averaged profiles of ion density $n_i(r)$, plasma potential $\phi(r)$, and electron temperature $T_e(r)$ against the uncertainty band from the PPPL long-time simulation~\cite{landmark_penning}, defined as the minimum and maximum of a rolling 300~$\mu$s average over 7000 windows.

The ion density profile of (a) peaks near the axis and decays monotonically to near-zero at the wall, broadly consistent with the benchmark band. A modest overprediction of both the on-axis and wall density is present, which may be attributable primarily to the cylindrical domain introducing geometric compression towards the axis that is absent in the Cartesian benchmark. The corners of the benchmark square further provide additional surface area for particle loss at oblique angles while simultaneously enclosing a larger plasma volume, which may act to produce a somewhat lower density than the inscribed cylinder at equivalent injection rates. 

The elevated plasma density also implies a reduction in the local Debye length. Consequently, the wall sheath thickness is reduced. This behaviour is evident in the steeper potential gradient observed near the wall in (b). Otherwise, however, the potential lies well within the benchmark region with about 0.1 V discrepancy on-axis. The more negative plasma potential in the bulk is also consistent with the higher plasma density; to maintain the ambipolar balance between electron and ion losses, the plasma develops a deeper potential well, increasing electron confinement and reducing radial electron transport to the wall. 

Finally, panel (c) gives the electron temperature. Temperature measurements are particularly susceptible to systematic errors in regions of low particle occupancy, as documented at length in the benchmark paper~\cite{landmark_penning}. The azimuthal averaging inherent to the quasi-3D decomposition---whereby the $m=0$ component accumulates contributions from all azimuthal positions simultaneously at every time step---provides enhanced statistical coverage relative to a Cartesian single-snapshot sample, tending to reduce the noise floor and improve agreement with the long-time benchmark reference. The lower temperature by about 1 eV observed near the wall is a direct consequence of the stronger sheath structure; energetic electrons are preferentially lost to the wall and lower-energy electrons are present in the near-wall population. Within the bulk however, the temperature falls acceptably within the benchmark range.

Taken together, the benchmark profiles are reproduced with good agreement within the given range, while the quantitative differences are attributable to geometry-dependent transport and the resulting changes in sheath structure that most likely arise due to a cylindrical instead of Cartesian formulation. 

{\color{black}Finally, to verify the stability of the benchmark, the conservation
properties are examined through the time evolution of the total energy and angular momentum, the latter being chosen as a diagnostic of the rotational dynamics that dominate the Penning discharge, shown in Figure~\ref{fig:conservation}(a) and Figure~\ref{fig:conservation}(b)
respectively. The total energy includes both the particle kinetic energy and
the electrostatic field energy, while the momentum is defined as the
axial component of the particle angular momentum, $\ell_z = \sum_{s,k} m_s r_k v_{\theta,k}$. 

Following the initial transient associated with the formation of the rotating
spoke, both the energy and angular momentum settle into a quasi-steady state
and fluctuate about a well-defined mean value. These fluctuations are an
intrinsic feature of the open, non-linear and chaotic PIC system (particles are injected, lost to boundaries etc.). As expected for a magnetised electron–ion plasma, the total energy is dominated by the electron kinetic energy contribution, with the ion kinetic and field energy components providing smaller but dynamically important contributions. The temporal modulation of the field energy primarily reflects the response to the evolution of the electron population associated with the rotating spoke structure. In particular, this periodic modulation is phase-locked to the spoke rotations, reflecting the physical exchange of energy and angular momentum within the plasma structure. The RMS fluctuation of the total energy about its mean value is approximately $10\%$, with peak excursions of approximately $\pm15\%$, while the corresponding
RMS angular momentum fluctuation is approximately $7\%$. Similar magnitude fluctuations were observed in the Penning simulations of Tyushev et al\cite{fluc}. Importantly, neither
quantity exhibits a secular drift or unbounded growth over the simulation
duration. The absence of systematic energy increase or momentum imbalance
thus confirms that the implementation does not introduce numerical heating, finite-grid
instabilities, or pathological conservation errors.}

{\color{black}A central motivation for the Fourier decomposition is the reduction in the effective dimensionality of the simulation while retaining the essential higher-dimensional physics. Where the benchmark requires solving a two-dimensional Poisson problem, the DECOMPIC implementation here reduces that to 13 one-dimensional problems (25 real-valued Poisson problems). At fixed particles-per-cell $N_{ppc}$, this reduced-order representation also permits a correspondingly smaller particle population. Table~\ref{tab:cpu_comparison} compares the approximate wall-clock runtimes and equivalent CPU-hours reported in the benchmark reference~\cite{landmark_penning} for the complete 500~$\mu$s runs of 16 codes, with the corresponding DECOMPIC result. The benchmark codes consumed between approximately 3500 and $\gtrsim 250000$~CPU-hours for the complete run. The fastest codes were GPU-accelerated implementations (LTP-PIC on $4\times$ GPUs, PICASO-G on $1\times$ GPU, and RUBPIC on $1\times$ GPU) and the most CPU-intensive were large multi-core CPU deployments (PANTERA on 768 cores, EDIPIC-2D on 256 cores). In comparison, the present, as-yet unoptimised Java implementation of DECOMPIC completes the benchmark in approximately 40~h using 16 CPU cores, corresponding to 640~CPU-hours. This is approximately 46 times lower than the median CPU-code cost of about 30000~CPU-hours among the benchmark participants. In terms of wall-clock time, DECOMPIC is also faster than several of the mixed CPU/GPU implementations. However, the pure-GPU PICASO-G implementation remains four times faster, completing the benchmark in approximately 10~h~\cite{landmark_penning}.}

{\color{black}This illustrative comparison should nevertheless be interpreted with caution, since differences in numerical algorithms, code implementations, parallelisation strategies, hardware architectures, and resource utilisation mean that the reported CPU-hour and wall-clock costs do not constitute a definitive comparison of performance under equivalent computational conditions. {\color{black}In addition, the Cartesian benchmark and cylindrical DECOMPIC configurations are not spatially equivalent representations; the corners of the Cartesian domain extend to larger effective radii than the corresponding cylindrical boundary, resulting in a larger represented spatial region and, at fixed $N_{ppc}$, a differently sized particle population. The 46$\times$ reduction in CPU-hours should therefore be understood as an exemplification of the overall computational benefit of the reduced-order formulation, rather than as a direct measure of speed-up at fixed problem size.}

It is important to focus on the immediate benefit of the present approach as the reduction in the number of degrees of freedom required to represent the plasma system. The field cost reduction arises from the dimensional (mesh) reduction enabled by the Fourier decomposition, together with the improved conditioning of the modal Poisson solves for $m \geq 1$. Since the particle-operations (deposition, trajectory integration, interpolation, etc.) cost per step scales with the total number of macroparticles and, for a given $N_{ppc}$ loading, with the total number of cells, the DECOMPIC approach also supports a substantially lower absolute particle count. These combined reductions in field and particle degrees of freedom place the total computational cost per step substantially (roughly one to two orders of magnitude) below that of the corresponding full 2D Cartesian representation, with the precise reduction depending on the relative cost of the field solve versus particle operations.} 

The savings scale generically with the degree to which the discharge retains azimuthal organisation and the number of modes $N_m$ required. Since DECOMPIC is currently an unoptimised implementation, the introduction of even modest additional parallelism strategies (for example, domain decomposition beyond the modal load-balancing currently implemented) would extend the performance advantage further on modern multi-core hardware. These results demonstrate that the Fourier decomposition provides a practically competitive and physically accurate {\color{black}alternative to the conventional 2D approach for simulating radial-azimuthal plasma devices}, without recourse to GPU acceleration or large-scale distributed-memory parallelism.

\section{Conclusions}

A {\color{black}reduced-order} Particle-in-Cell algorithm, DECOMPIC, has been developed and validated for kinetic simulation of azimuthal instabilities and cross-field transport in nominally axisymmetric magnetised plasmas. The method rests on a single structural observation: Fourier orthogonality in the azimuthal direction decouples the three-dimensional Poisson equation exactly into a family of independent two-dimensional elliptic problems on the meridional plane, one per retained azimuthal mode. No approximation beyond spectral truncation is introduced; the full three-dimensional phase space of the macroparticle ensemble is preserved throughout. The resulting modal solves are better conditioned than the axisymmetric base problem by virtue of the mode-dependent centrifugal diagonal increment, which also makes Jacobi preconditioning both natural and sufficient. These properties distinguish the method from competing reduced-cost approaches ---separability-ansatz, sparse-grid, and machine-learning schemes --- which generally introduce uncontrolled or empirically adjusted approximations into the field solve, whose impact depends on the application.

Validation against the diocotron instability of a hollow electron annulus demonstrated systematic quantitative agreement with analytic linear theory across three geometrically distinct configurations. Growth rates extracted by least-squares regression over the exponential phase agree with the closed-form prediction to within 7\%, and the non-linear vortex dynamics — roll-up, discrete vortex formation, and merger — are qualitatively reproduced in a manner consistent with well-documented behaviour of this instability. Convergence studies established that the dominant numerical error is statistical shot noise, following the expected {\color{black}$\sqrt{N_{ppc}}^{-1}$} scaling and becoming negligible at $N_{ppc}>200$ particles per cell. The extracted growth rate is robust to the retained mode number $N_m$ across the full range tested, confirming that the linear physics is captured correctly once the dominant mode is resolved.

The Landmark Penning discharge benchmark demonstrated the practical computational advantage of the approach. DECOMPIC reproduces the rotating-spoke frequency (43.05 kHz against the reference value of 43.2 kHz), the modal energy hierarchy, and the time-averaged radial profiles of ion density, plasma potential, and electron temperature in quantitative agreement with the long-time reference simulation, at a cost of approximately 640 CPU-hours on a standard multi-core desktop — a factor of 46 below the median benchmark participant and competitive with GPU-accelerated codes, without recourse to distributed-memory parallelism or hardware accelerators. Residual discrepancies in the density and sheath profiles are attributable to the cylindrical rather than Cartesian geometry of the present formulation and represent a physically consistent difference rather than a numerical artefact. 

{\color{black}The method has therefore demonstrated that low-order spectral representations of ignorable coordinates can recover essential non-axisymmetric kinetic plasma dynamics at a computational cost intermediate between conventional reduced-dimensionality models and equivalent full-dimensionality PIC simulation.}

However, several limitations of the current implementation warrant acknowledgement. The formulation is currently electrostatic: self-generated magnetic fields are neglected on the basis of low plasma beta, and the method as presented is not appropriate for regimes in which magnetic instabilities are dynamically significant. {\color{black}More fundamentally, the validation presented here --- the diocotron instability and the Landmark Penning discharge --- involves no significant axial dynamics; the Penning case in particular was run with a single axial cell (to conform to the benchmark). The reduction of the three-dimensional field solve to a family of two-dimensional problems is an exact consequence of the governing equations and therefore holds generally, but the practical computational advantage for configurations with substantial axial structure has not yet been demonstrated and is the specific target of a planned follow-on study.} The mode truncation at $N_m$ may also introduce Gibbs-type spectral aliasing for instabilities with fine azimuthal structure; in practice the rapid spectral decay of the modal energy spectrum for smooth cylindrical plasmas ensures a modest $N_m$ suffices, but this must be verified for each new configuration. Indeed, the efficiency of the approach depends upon rapid convergence of the azimuthal spectrum. Problems characterised by strongly broadband non-axisymmetric structure may require a larger number of retained modes, reducing the computational advantage. Finally, the current Java implementation, while already competitive, has also not been subjected to the low-level optimisation that would further extend the performance margin.

Looking ahead, the most significant extension is the self-consistent computation of quasilinear transport coefficients from the resolved modal fields. The turbulent flux and anomalous diffusivity follow directly from the cross-correlation of the modal density and electric-field amplitudes and can be accumulated at negligible additional cost during the field reconstruction step, removing the need for the empirical closure assumptions that presently limit predictive fidelity in plasma propulsion and other plasma device modelling. Extension to electromagnetic operation via a modal decomposition of the vector potential, and to non-periodic axial boundaries representative of open plasma devices, are further natural directions. {\color{black} Regarding the methodology, {\color{black}improvement may be found} in parallelisation domain decomposition beyond the implemented modal parallelism, $m>0$ axis boundary condition treatment (including higher-order discretisation or regularity-aware reconstruction), non-uniform azimuthal weighting of macroparticles into the modal subspaces, and the application of typical sub-cycling or scaling techniques within the modal framework}. {\color{black}Together, these developments would represent a necessary step toward establishing DECOMPIC as a reduced-cost kinetic framework capable of helping to close the gap between computationally intractable full-3D simulation and the physically limited reduced-dimensionality models on which plasma propulsion engineering currently relies.} 

\section*{Data Availability Statement}
The data that supports the findings of this study are available from the corresponding author upon reasonable request.


%
%

%


\section*{References}
\bibliography{bibliography}
\newpage
\appendix
{\color{black}
\section{Effective Plasma-Frequency and Debye-Length Scales of the Fourier Modes}
\label{sec:appendixA}

The characteristic plasma-frequency and Debye-length scales associated with the Fourier modes are derived here directly from the modal representation. The purpose is to distinguish the equilibrium density, which determines the collective response of the plasma, from the amplitude of an individual Fourier coefficient, which represents a perturbation about that equilibrium.

The electron density, velocity, electric field, and electrostatic potential are represented by Fourier series consistent with Eqs.~(\ref{eq:phi_series})--(\ref{eq:rho_series}):
\begin{align}
&n_e(r,\theta,z,t)
=
\sum_{m\in\mathbb{Z}}
\hat n_{e,m}(r,z,t)e^{-im\theta},\\
&
\mathbf{u}_e(r,\theta,z,t)
=
\sum_{m\in\mathbb{Z}}
\hat{\mathbf{u}}_{e,m}(r,z,t)e^{-im\theta},\\
&\mathbf{E}(r,\theta,z,t)
=
\sum_{m\in\mathbb{Z}}
\hat{\mathbf{E}}_m(r,z,t)e^{-im\theta},\\
&
\phi(r,\theta,z,t)
=
\sum_{m\in\mathbb{Z}}
\hat{\phi}_m(r,z,t)e^{-im\theta}.
\label{eq:app_modal_expansions}
\end{align}
Consider an electrically neutral cold plasma in equilibrium of form
\begin{equation}
\hat n_{e,0}=n_{e,0},
\qquad
\hat{\mathbf{u}}_{e,0}=0,
\qquad
\hat{\mathbf{E}}_0=0,
\end{equation}
while the complex coefficients $\hat n_{e,m}$, $\hat{\mathbf{u}}_{e,m}$, $\hat{\mathbf{E}}_m$, and $\hat{\phi}_m$ for $m\neq0$ represent perturbations about this equilibrium. The following derivations consider the linear response of these perturbations.

\subsection{Effective modal plasma frequency}
\label{app:modal_plasma_frequency}

Consider the cold electron continuity and momentum equations together with Poisson's equation,
\begin{align}
&\frac{\partial n_e}{\partial t}
+
\nabla\cdot(n_e\mathbf{u}_e)
=0,
\\
&m_e\frac{\partial\mathbf{u}_e}{\partial t}
=
-e\mathbf{E},\\
&\nabla\cdot\mathbf{E}
=
-\frac{e}{\varepsilon_0}
\left(n_e-n_i\right).
\label{eq:app_fluid_equations}
\end{align}
Substituting the Fourier representations of Eqs.~(A1-\ref{eq:app_modal_expansions}) and retaining terms to first order in the perturbations gives
\begin{align}
&\frac{\partial \hat n_{e,m}}{\partial t}
+
\nabla_{rz}\cdot
\left(
n_{e,0}\hat{\mathbf{u}}_{e,m}
\right)
+
\frac{im}{r}n_{e,0}\hat u_{\theta,m}
=0,
\label{eq:app_modal_continuity}
\\
&m_e\frac{\partial\hat{\mathbf{u}}_{e,m}}{\partial t}
=
-e\hat{\mathbf{E}}_m,
\label{eq:app_modal_momentum}
\\
&\frac{1}{r}\frac{\partial}{\partial r}
\left(r\hat E_{r,m}\right)
+
\frac{\partial\hat E_{z,m}}{\partial z}
-
\frac{im}{r}\hat E_{\theta,m}
=
-\frac{e}{\varepsilon_0}\hat n_{e,m}.
\label{eq:app_modal_poisson}
\end{align}
Here the azimuthal derivative has been replaced by $\partial_\theta\rightarrow-im$ following the convention of Eq.~(\ref{eq:phi_series}). The non-linear terms in the perturbations, such as products of two non-zero modal quantities, are omitted under linearisation.

For a locally homogeneous equilibrium, $n_{e,0}$ may be treated as constant over the perturbation scale. Taking the time derivative of Eq.~(\ref{eq:app_modal_continuity}) and using Eq.~(\ref{eq:app_modal_momentum}) gives
\begin{equation}
\frac{\partial^2\hat n_{e,m}}{\partial t^2}
=
\frac{n_{e,0}e}{m_e}
\left[
\frac{1}{r}\frac{\partial}{\partial r}
\left(r\hat E_{r,m}\right)
+
\frac{\partial\hat E_{z,m}}{\partial z}
-
\frac{im}{r}\hat E_{\theta,m}
\right].
\end{equation}
Using Eq.~(\ref{eq:app_modal_poisson}) then gives
\begin{equation}
\frac{\partial^2\hat n_{e,m}}{\partial t^2}
+
\frac{n_{e,0}e^2}
{m_e\varepsilon_0}
\hat n_{e,m}
=
0.
\label{eq:app_modal_plasma_oscillation}
\end{equation}
Defining an effective modal plasma frequency through
\begin{equation}
\frac{\partial^2\hat n_{e,m}}{\partial t^2}
+
\omega_{pe,m}^2\hat n_{e,m}
=
0,
\end{equation}
therefore gives
\begin{equation}
\omega_{pe,m}
=
\sqrt{
\frac{n_{e,0}e^2}
{m_e\varepsilon_0}
}.
\label{eq:app_modal_omega}
\end{equation}
Thus,
\begin{equation}
\boxed{
\omega_{pe,m}
=
\omega_{pe,0}
=
\sqrt{
\frac{n_{e,0}e^2}
{m_e\varepsilon_0}
}.
}
\label{eq:app_modal_omega_equal}
\end{equation}

The modal amplitude $\hat n_{e,m}$ therefore determines the amplitude of the perturbation but does not determine its characteristic plasma frequency. The equilibrium density remains the coefficient determining the linear collective restoring response for every mode.

\subsection{Effective modal Debye length}
\label{app:modal_debye}

The corresponding screening scale can be obtained from the low-frequency, isothermal electron response. Neglecting electron inertia, the electron momentum equation is

\begin{equation}
0
=
-e\mathbf{E}
-
\frac{k_BT_e}{n_e}\nabla n_e.
\label{eq:app_debye_force_balance}
\end{equation}
Substituting the Fourier representations of Eq.~(\ref{eq:app_modal_expansions}), using $\mathbf{E}=-\nabla\phi$, and linearising about the equilibrium gives the modal density response
\begin{equation}
\hat n_{e,m}
=
\frac{n_{e,0}e}
{k_BT_e}
\hat\phi_m.
\label{eq:app_modal_boltzmann}
\end{equation}

The modal Poisson equation follows from Eq.~(\ref{eq:modal_poisson}) and considering purely the electron perturbation,
\begin{equation}
\mathcal{L}_m\hat\phi_m
=
\frac{e}{\varepsilon_0}\hat n_{e,m}.
\end{equation}
Substitution of Eq.~(\ref{eq:app_modal_boltzmann}) gives
\begin{equation}
\mathcal{L}_m\hat\phi_m
-
\frac{n_{e,0}e^2}
{\varepsilon_0 k_BT_e}
\hat\phi_m
=
0.
\label{eq:app_modal_screening}
\end{equation}
Defining the effective modal Debye length by
\begin{equation}
\mathcal{L}_m\hat\phi_m
-
\frac{1}
{\lambda_{D,m}^2}
\hat\phi_m
=
0.,
\label{eq:app_modal_debye_definition}
\end{equation}
gives
\begin{equation}
\lambda_{D,m}
=
\sqrt{
\frac{\varepsilon_0 k_BT_e}
{n_{e,0}e^2}
}.
\label{eq:app_modal_debye}
\end{equation}
Hence,
\begin{equation}
\boxed{
\lambda_{D,m}
=
\lambda_{D,0}
=
\sqrt{
\frac{\varepsilon_0 k_BT_e}
{n_{e,0}e^2}
}.
}
\label{eq:app_modal_debye_equal}
\end{equation}

Thus, the Fourier decomposition modifies the spatial structure of the response through the azimuthal wavenumber $m/r$ being present in the modal field operator $\mathcal{L}_m$, but it does not modify the Debye screening. The modal density amplitude $\hat n_{e,m}$ likewise does not define a separate screening length because it is the perturbation responding to the equilibrium plasma, rather than an equilibrium density itself. 

It is the equilibrium density that determines the characteristic collective plasma-frequency and Debye-length scales.}

{\color{black}

\section{Method of Manufactured Solutions Verification}
\label{app:mms}

\begin{figure*}[!t]
\centering
\includegraphics[width=0.85\textwidth]{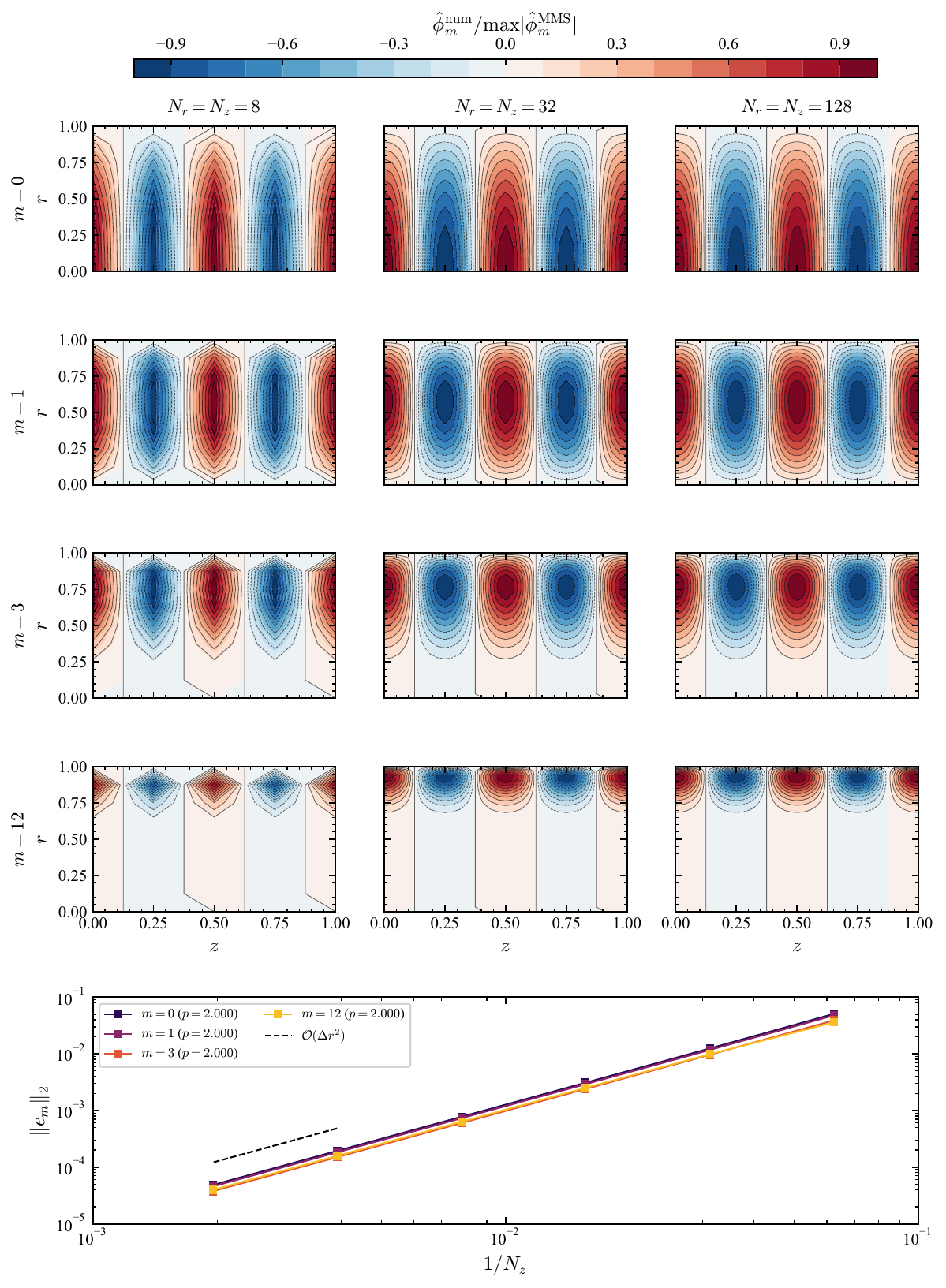}
\caption{Method of Manufactured Solutions (MMS) verification of the second-order discretisation error in the modal Poisson solve.}
\label{fig:mms}
\end{figure*}

The spatial accuracy of the modal Poisson solver was assessed using the Method of
Manufactured Solutions (MMS). The MMS provides a rigorous verification of the
numerical implementation by prescribing an analytical solution and constructing
the corresponding source term such that the governing equation is satisfied
exactly. Since the exact solution is known everywhere within the computational
domain, the numerical error can be evaluated directly, allowing the observed
order of convergence of the discretisation to be quantified independently of
any physical benchmark or modelling assumptions.

To exercise the full two-dimensional discretisation, including the axial
second-derivative operator, the manufactured solution was chosen to possess both
radial and axial variation,

\begin{equation}
\hat{\phi}_m(r,z)
=
r^m
\left(R_{\max}^2-r^2\right)
\cos(k_z z),
\label{eq:mms_phi}
\end{equation}

where

\begin{equation}
k_z=\frac{2\pi N_{\mathrm{waves}}}{L_z},
\end{equation}

with $R_{\max}=1$, $L_z=1$, and $N_{\mathrm{waves}}=2$ for the present study.
The polynomial radial dependence ensures regular behaviour at the axis while
simultaneously satisfying a homogeneous Dirichlet condition at the outer
conducting wall. The cosine dependence satisfies the periodic axial boundary
conditions and introduces a non-trivial axial variation, ensuring that the
complete two-dimensional operator is verified.

The source term corresponding to the manufactured solution was obtained by
substituting Eq.~(\eqref{eq:mms_phi}) into the governing equation,

\begin{equation}
\hat{\rho}_m(r,z)
=
-
\varepsilon_0
\mathcal{L}_m\hat{\phi}_m,
\label{eq:mms_rho_operator}
\end{equation}

which yields

\begin{equation}
\hat{\rho}_m(r,z)
=
\varepsilon_0
\,r^m
\left[
4(m+1)
+
k_z^2\left(R_{\max}^2-r^2\right)
\right]
\cos(k_z z).
\label{eq:mms_rho}
\end{equation}

Consequently, the only remaining source of error arises from the spatial discretisation of the
Poisson operator.

At the axis, the axisymmetric mode ($m=0$) satisfies a
homogeneous Neumann condition, while all higher-order modes satisfy homogeneous
Dirichlet conditions arising from regularity. A homogeneous Dirichlet condition
is imposed at the outer conducting wall and periodic boundary conditions are
applied in the axial direction.

\begin{equation}
\begin{aligned}
\frac{\partial\hat\phi_m}{\partial r}(0,z)&=0,
&&m=0,\\
\hat\phi_m(0,z)&=0,
&&m>0,\\
\hat\phi_m(R_{\max},z)&=0,\\
\hat\phi_m(r,L_z)&=\hat\phi_m(r,0).
\end{aligned}
\end{equation}

The numerical parameters used throughout the verification are listed in
Table~\ref{tab:mms_settings}.

\begin{table}[h]
\centering
\caption{Parameters for the MMS verification.}
\label{tab:mms_settings}
\begin{tabular}{ll}
\hline
Parameter & Value \\
\hline
$R_{\max}$ & 1 \\
$L_z$ & 1 \\
$\varepsilon_0$ & 1 \\
$N_{\mathrm{waves}}$ &
2 \\
$N_r=N_z$ &
[8,16,32,64,128,256,512] \\
$m$ &
[0,1,3,12] \\
\hline
\multicolumn{2}{c}{\textit{Multigrid solver parameters}}\\
\hline
$\|L_m\hat\phi_m-b_m\|_2/\|b_m\|_2 $ &
$10^{-8}$ \\
$\omega$ & 1.4 \\
Pre/Post Smoothing Sweeps & 3 \\
$\beta$ &
1.2 \\
Min. Coarse Grid $N_r=N_z$ & 8 \\
\hline
\end{tabular}
\end{table}

The multigrid parameters were selected to ensure that the algebraic error
introduced by the iterative linear solver remained negligible compared with the
spatial discretisation error. In particular, the V-cycle iterations were
terminated when the relative residual was reduced below $10^{-8}$. This tolerance is several orders of magnitude smaller than the discretisation errors observed in the MMS study, which remain
within the range $\mathcal{O}(10^{-5})$--$\mathcal{O}(10^{-1})$ over the tested grid resolutions. Consequently, the measured convergence rate is controlled by the finite-volume
discretisation rather than by the convergence of the multigrid.
The MMS calculations employ a radial stretching of $\beta=1.2$, which ensures mode-dependent mesh criterion are satisfied for up to $m=12$.

The numerical error was then quantified using the relative weighted
$L_2$ norm,

\begin{equation}
\frac{\|e_m\|_2}{\|\hat\phi_m\|_2}
=
\frac{
\left(
\sum (\hat\phi_m^{\mathrm{num}}-\hat\phi_m^{\mathrm{MMS}})^2
\,r\,\Delta r\,\Delta z
\right)^{1/2}}
{\left(
\sum (\hat\phi_m^{{\mathrm{MMS}}})^2
\,r\,\Delta r\,\Delta z
\right)^{1/2}},
\end{equation}

where the cylindrical weighting by $r$ is consistent with the conservative form of the
modal Poisson equation. The observed convergence order, $p$, follows from a least-squares linear
fit to the error data on a log-log scale,

\begin{equation}
\log\left(\|\mathbf{e}_m\|_2\right)
=
\log(C_m)+p\log(h),
\label{eq:mms_convergence_fit}
\end{equation}

where $h=1/N_z$ is defined as the characteristic mesh spacing and $C_m$ is a constant.

Figure~\ref{fig:mms} presents the results of the verification. The upper panels
show the numerical solution obtained on grids with $N_z=N_r=8$, $32$ and $128$ for a
selection of representative azimuthal modes. As the mesh is refined, the
numerical solution exhibits progressively smoother spatial variation and the
grid-scale discretisation error rapidly diminishes. By $N_z=N_r=128$, no visible
numerical artefacts remain and the solution is effectively grid independent.

The lower panel presents the convergence study for the representative modes
$m=0$, $1$, $3$ and $12$. In every case the error decreases with the square of
the mesh spacing, producing measured convergence rates of approximately
$p=2$. This confirms that the finite-volume implementation achieves the
expected spatial accuracy over a wide range of azimuthal modes. Furthermore, the
nearly identical convergence behaviour observed for low and high mode numbers
demonstrates that the accuracy of the discretisation is maintained irrespective
of the strength of the centrifugal term. The MMS verification therefore provides
strong evidence that the solver is correctly implemented and that the observed
errors are dominated by the expected second-order discretisation error. Furthermore, for characteristic grid resolutions representative of practical simulations, the measured
$L_2$ errors remain sufficiently small $<O(10^{-4})-O(10^{-3})$.

Although all modes exhibit the expected second-order convergence rate, the magnitude of the $L_2$ error varies between modes. The reduction in error from $m=0$ to $m=1$ and $m=3$ is attributed primarily to differences in the radial spatial structure of the manufactured solutions. Higher-order modes also introduce an additional diagonal contribution to the discrete operator, which has been {\color{black}established} to improve the conditioning of the linear system. However, the $m=12$ case exhibits a slightly larger error than $m=3$. This may be due to increased radial localisation and stronger gradients associated with this higher mode.}

\end{document}